\documentclass[twocolumn,aps,pre,showpacs,preprintnumbers,amsmath,amssymb, superscriptaddress]{revtex4}

\usepackage{graphicx}
\usepackage{dcolumn}
\usepackage{bm}

\begin{document}


\title{Time Scales for Rounding of Rocks through Stochastic Chipping} 


\author{D. J. Priour, Jr}
\affiliation{Department of Physics \& Astronomy, Youngstown State University, Youngstown, OH 44555, USA}


\date{\today}

\begin{abstract}
For three dimensional geometries, we consider stones (modeled as convex polyhedra)
subject to weathering with planar slices of random orientation and depth successively removing material,
ultimately yielding smooth and round (i.e. spherical) shapes.  An exponentially decaying acceptance probability
in the area exposed by a prospective slice provides a stochastically driven physical basis for the removal of material
in fracture events.  With a variety of quantitative measures, in steady state we find a
power law decay of deviations in a toughness parameter $\gamma$ from a perfect spherical shape.
We examine the time evolution of shapes for stones initially in the form of cubes as well as irregular fragments
created by cleaving a regular solid many times along random fracture planes.   In the case of the former, we
find two sets of second order structural phase transitions with the usual hallmarks of critical behavior.
The first involves the simultaneous loss of facets inherited from the parent solid, while the
second transition involves a shift to a spherical profile.
Nevertheless, for mono-dispersed cohorts of irregular solids, the loss of primordial
facets is not simultaneous but occurs in stages.
In the case of initially irregular stones, disorder obscures individual structural transitions, and relevant
observables are smooth with respect to time. More broadly, we find that times for the achievement 
of salient structural milestones scale quadratically in $\gamma$.
We use the universal dependence of variables on the fraction of the original volume remaining to calculate
time dependent variables for a variety of erosion scenarios 
with results from a single weathering scheme such as the case in which the
fracture acceptance probability depends on the relative area of the prospective new face.
In this manner, We calculate time scales of interest and also obtain closed form approximate
expressions which bound direct simulation results from above.
\end{abstract}

\maketitle
\section{Introduction}

Various types of mechanisms act to weather or transform the shapes of rocks by eroding away 
their volume.  Many of these processes, such as collision induced fragmentation, are 
inherently stochastic.  Salient questions related to the emerging shapes as more and more 
volume is removed by random interactions with neighboring stones include the degree to which
rock surfaces become smooth on a macroscopic scale, as well as the extent to which the 
accumulating random fractures yield spherical profiles,
and the rate at which  mass is removed through weathering.
We investigate these concerns  with large scale Monte Carlo
simulations in which primordial convex polyhedral rocks (i.e. proto-clasts) with a small number 
of faces are subject to a sequence of planar fractures which successively carve away 
material from the parent stones.  

With a combination of quantitative measures of the deviation     
from a perfect spherical shape, the overall oblateness and prolateness, and smoothness in terms 
of how evenly the total surface area is distributed over the individual facets, we characterize 
the evolution of the shapes of ensembles of 
rocks as stochastic fractures accumulate and volume is worn away. To statistical Monte Carlo statistical error, we follow the shape 
trajectories of at least 1000 stones, which accumulate as many as $2 \times 10^{6}$ planar slices, yielding 
polyhedra with up to 3,600 facets on average. 

Using a two-pronged approach, we find that stones subject to stochastically driven collisional weathering ultimately
converge to spherical shapes.  On the one hand, we consider a steady state scenario where observables (apart from 
the volume, which decreases monotonically in time) stabilize at constant values and thus in a sense equilibrate.  
By tuning a parameter $\gamma$ controlling the coarseness of the shapes at equilibrium (i.e. the mean relative area 
$\langle \Delta A \rangle /A_{\Sigma}$ of newly formed facets), we find that the equilibrium mean number $\langle n \rangle$ of planar faces
scales as $\gamma$ while quantitative measures of the deviation from spherical shapes and/or a  
perfectly smooth surface tend systematically to zero with increasing $\gamma$.  On the other hand, 
we also calculate the time dependence (i.e. with the number of sustained slices $N_{\mathrm{sust}}$ serving as a proxy for time) 
of observables in the scenario in which the latter eventually equilibrate.  
We use the latter to obtain corresponding quantities for a broad range of non-steady state
scenarios, a technique we validate with direct Monte Carlo simulation in the context of the 
fixed velocity scheme, envisaged for rocks impelled by a current at a mean constant speed independent of 
their mass.  In this way, we glean quantitative results for time scales for the attainment of milestones such as the 
removal of a specific fraction of the primordial volume, finding that such times vary depending on the specific erosion scenario under 
consideration but do not diverge, a point we underscore with approximate closed form expressions also yielding finite times   
while serving as upper bounds for simulation results.

Taking into consideration that in practice cohorts of primordial stones prior to weathering may be structurally diverse,
we consider protoclasts in the form of irregularly shaped polyhedra to account for aggressive fragmentation events prior to the collision induced 
weathering.  To mimic structures resulting from dramatic fragmentation processes, initial cohorts of structurally disordered shapes are generated with 
a sequence of random fracturing planes, invariably accepted independent of 
the orientation and depth of the prospective slice.  With protoclast cohorts generated in this manner, we find relevant observables to vary smoothly with time. 
As companion calculations, we also examine structurally homogeneous primordial stones, including regular cubes,  
which ultimately are smoothed into spherical shapes just as occurs in the case of irregular protoclasts.  However 
in contrast to structurally disordered stones, observables in the case of cubes exhibit two continuous phase 
transitions with all of the hallmarks of a second order phase transition, including power law scaling for 
singularities in relevant observables as we show explicitly in this work. 
The first transition is a structural transformation in which the primordial cube facets are sheared away, while 
in a subsequent transition the stones derived from regular solids revert to spherical shapes with concomitant singularities in 
quantities sensitive to deviations from a perfect spherical profile.
A structural phase transition of the former type has been proposed in previous studies based on laboratory experiments and 
numerical studies~\cite{Domokos2,Domokos3}.  We apply the tools of finite size scaling to locate the transition and calculate associated critical exponents. 

In addition to regular shapes, we consider a homogeneous ensemble of geometrically irregular
solids finding asynchronous transitions for the erosion of primordial facets instead of a single structural 
transition for the solid as a whole, though again we find a single second 
order transition to a round spherical shape at a later time after all of the primordial facets have been removed.
  Thus, only in the case of highly symmetric shapes is there a simultaneous 
elimination of the facets inherited from the parent solid.

In spite of the stochastic nature of the 
erosion scheme, we find in the regime that fragments cleaved away through random fracture events are small 
in volume relative to that of the stone as a whole that shapes evolve deterministically,  
with reproducible changes  on scales small in comparison with the overall size of the stone but large relative to 
freshly exposed faces.
As in a recent study, we find a universal dependence of observables on the remaining volume fraction~\cite{Domokos1}.  We take advantage 
of this characteristic to calculate time dependent observables for a variety of erosion schemes using results from the scenario 
of fixed relative area for typical fresh facets created by fracture events.

Aside from measures of deviation from a perfectly spherical shape, we also examine more specific macroscopic
structural characteristics, including the degree to which stones are oblate or prolate as volume is 
chipped away.  While both the former and the latter tend to zero as rocks are rounded, 
the oblateness measure is non-monotonic in time, initially rising and then peaking after approximately 50\% of the 
original volume has been removed, and declining and tending to zero thereafter.

As a novel feature of the calculations in this work, no restrictions are imposed as to how many vertices and facets are 
truncated or removed by a sustained slice, and our calculation is compatible with the elimination of 
an arbitrary number of vertices per fracture.  For a plausible 
physical basis for fracture events, measures such as the 
area of a newly exposed face are considered to determine if a candidate fracture plane is 
accepted.  To set the scale of new facet areas, we implement 
an energy based criterion for accepting prospective slices, taking the 
kinetic energy input needed to cleave away a slice to be proportional to the area of the new face.  
In this vein, in the spirit of statistical mechanical treatments, we assume the fracture probability to    
be exponentially suppressed in the exposed area $\Delta A$, and given by 
$P(\Delta A) = e^{-\gamma \Delta A/\theta}$ where $\gamma$ is a constant related to the 
toughness of the material comprising the rocks and $\theta(V)$ is a volume  
dependent measure of the mean 
kinetic energy or ``temperature'', with volume serving as a proxy for the mass in the 
case of stones of uniform composition as assumed in this manuscript.  
In addition, since we consider initial cohorts of stones to be mono-dispersed  with respect to mass
(though structurally diverse in the case of irregular protoclasts), we operate in terms of the relative volume 
$\tilde{v} \equiv V/V_{0}$ with $V_{0}$ the volume prior to the erosion process.
With analytical arguments and direct simulation results, we show that characteristic times scale 
quadratically in the toughness parameter $\gamma$.

In this work to be definite we assume a power law dependence $\theta = \theta_{0} \tilde{v}^{\alpha}$ (i.e. with $\alpha > 0$), 
reflecting kinetic energy dependence on volume as a rock is borne along (e.g. by water currents 
in a river or stream as collisions chip away material).  Defining $\gamma^{'} \equiv \gamma/\theta_{0}$ for 
the sake of brevity, the 
acceptance probability for a prospective fracture plane is $P(\Delta A) = e^{-\gamma^{'} \Delta A \tilde{v}^{-\alpha} }$.  Assuming 
$\gamma^{'} \langle \Delta A \rangle \tilde{v}^{-\alpha} \sim 1$ for the mean area $\langle \Delta A \rangle$ of a new facet,  
one would argue on dimensional grounds that the mean number of faces is 
$\langle n \rangle \sim A_{\Sigma}/\langle \Delta A \rangle = \gamma^{'} \tilde{v}^{-\alpha} A_{\Sigma}$, where $A_{\Sigma}$ is the total 
polyhedron surface area.
For the 3D case, taking the total area to scale as $\tilde{v}^{2/3}$, $\alpha_{c}$ marks a boundary between stones which 
in principle could become smoother over time ($\alpha > \alpha_{c}$) and shapes which become coarser ($\alpha < \alpha_{c}$)
as fractures accumulate; whereas relative areas  $\langle \Delta A \rangle /A_{\Sigma}$ of new facets decrease for the former, newly
exposed facets encompass an increasingly large share of the surface in the case of the latter.
The boundary case $\alpha = \alpha_{c}$ thus offers the possibility for the 
attainment of a steady state for the mean relative area of new faces, as well as other variables of interest.    

Another merit of obtaining steady state or relative area results is the fact that the universal dependence 
of observables on volume fraction permits one to map remaining stone volumes onto time for
a scheme distinct from the relative area case, an technique bolstered with results from
large-scale Monte Carlo simulations.  We show in this manuscript that 
both the former and latter are identical up to random Monte Carlo statistical error.
Time scales obtained in this manner are then used to validate a closed form   
expression serving as an approximation and strict upper bound for times associated with 
the chipping away of a specific volume fraction or the structural 
phase transition for cohorts of regular protoclasts in which primordial facets disappear.  

In Section II, we discuss principal quantitative measures of deviation from sphericity as well as key 
methodological elements.  Section III contains results for rocks in the case of the steady state 
scenario, while Section IV broadens the discussion to non-steady state situations.  In Section V, we calculate
time scales for a range of erosion schemes, with a closed form expression obtained as an upper bound for 
time scales for salient structural milestones.
Conclusions are found in Section VI.

\section{Quantitative Measures and Methodology}

Two salient quantities which permit one to address in a direct manner the degree to which a shape departs from a 
perfect spherical profile include the configuration averaged sphericity $\phi_{3}$ and the ratio $r^{\mathrm{min}}_{\mathrm{max}}$ 
of the minimum and maximum distance of
surface points for the center of mass.  Both measures distinguish among a spherical shape and 
a stone which is not perfectly round.  The sphericity benefits from the fact that 
surface area to volume ratio of a solid attains a global minimum for spheres, and is given by 
$\phi_{3} \equiv [6 \pi^{\frac{1}{2}} V]^{\frac{2}{3}}/A_{\Sigma}$~\cite{Wadell}, where $A_{\Sigma}$ is the surface area and $V$ the volume.
Since $\phi_{3} < 1$ (except in the case of a perfect sphere where $\phi_{3} = 1$), the complement $1 - \phi_{3}$ 
serves as a measure of the departure from a perfectly spherical shape.  An alternative measure, 
sensitive to local features,
is $r^{\mathrm{min}}_{\mathrm{max}} \equiv d_{\mathrm{min}}/d_{\mathrm{max}}$, the ratio of the minimum
and maximum distances, respectively, of points on the rock surface from its center of mass.
In the context of our calculations involving faceted objects, $d_{\mathrm{min}}$ is the distance to 
the closest planar facet and  $d_{\mathrm{max}}$ is the distance to the farthest vertex.
Again, $r^{\mathrm{min}}_{\mathrm{max}}$ peaks at unity only in the case of a perfect sphere, with 
$1 - r^{\mathrm{min}}_{\mathrm{max}}$ serving as a measure of the deviation from a perfect spherical shape.

Related to whether there is a systematic convergence to a spherical shape is 
if erosion through the stochastic chipping mechanism yields a stone with a smooth surface.  Although sphericity 
implies smoothness, the converse is not true.  In fact, in this work we find an example of a perfectly smooth non-spherical 
shape at the phase transition for structurally homogeneous cohorts where primordial facets disappear.  
Hence, as a measure of smoothness not anchored to a specific 
geometry, we use the Inverse Participation Ratio (IPR), where $\textrm{IPR} = \langle A_{\Sigma}^{-2} \sum_{i=1}^{N} A_{i}^{2} \rangle$
with $A_{\Sigma}$ being the total surface area, $A_{i}$ the area of the $i$th of $N$ facets, and angular brackets indicated a configurational 
average.  The IPR tends to zero with increasing $N$ for an even distribution of the area of the polyhedron faces (i.e. for
smooth shapes),
but converges to a finite value if a small number of facets contain a macroscopic fraction of $A_{\Sigma}$ (i.e for rough stones),  
and thus serves as a quantitative measure of smoothness.  Appealing to the IPR in this manner is analogous to its 
usage in the context of charge transport discussions to distinguish among extended and localized carrier states~\cite{Wegner}.

In this work, we simulate collisional weathering with large scale Monte Carlo simulations involving sequences of randomly oriented 
planar fractures, where rocks begin as regular or irregular protoclasts with a small number of 
faces [i.e. 6 and a mean of 9.03(1) sides for cubic and irregular polyhedra respectively], ultimately yielding polyhedra
with as many as 3,600 facets.  Although there are qualitative differences for the time dependence of observables in the cases of 
cube shaped and irregular protoclasts (e.g. the structural phase transition for cube shaped protoclasts not seen in ensembles 
of irregular shapes), both cohorts ultimately converge to spherical shapes in the long time regime.  Moreover,
we obtain an approximate analytical expression for time scales for salient stages in the attainment of a perfectly spherical shape.  
The fact that these times are finite,  bounding corresponding numerical results from above is compatible with the eventual conversion of 
rocks into spherical shapes for a wide range of erosion scenarios.

In stochastic chipping sequences, each sustained slice exposes a new face while truncating one or more vertices.  The 
likelihood of accepting a prospective slice hinges solely on the area $\Delta A$ and volume of the parent 
solid with no restriction on the number of truncated vertices, raising the possibility of the elimination of 
an entire face or faces if each of the constituent vertices are cleaved away.  For the sake of an efficient 
implementation, updates of lists of vertices and planes making up the polyhedron include the addition of new 
features as well as the deletion of those eliminated by the most recent fracture event.
Indeed, in the steady state scenario, as many vertices and facets are removed on average as accumulate 
with each sustained slice.  Previous studies in two and three dimensional geometries have involved the truncation of a finite 
number of vertices or facets per fracture event~\cite{Durian2,Krapivsky,Domokos2,Domokos1}. A calculation involving two dimensional shapes 
allowed for the removal of an arbitrary number of vertices~\cite{Durian1}  To our knowledge, our calculations are the first to 
examine erosion phenomena due to a sequence of planar fractures allowing for the elimination of an arbitrary number 
of vertices and/or facets for three dimensional geometries. 

To facilitate the examination of polyhedra with a large number of faces, a variety of measures are 
employed to optimize the efficiency of locating vertices of prospective facets.  While one could 
in principle examine all possible intersections of a fracture plane with the faces comprising the 
stone, in general only a small subset of vertices identified in this manner populate the new face,  
comprising a share of the total which diminishes as the number of planar faces increases.
We avoid considering spurious vertices by only examining faces which contain vertices both above 
and below the prospective fracture plane, as facets not meeting this condition are either sheared away
altogether or are not truncated by slice at all and hence do not yield any new constituent vertices.
As a further constraint, we only seek intersections of the slicing plane with pairs of faces which
have an edge in common.  Finally, we avoid computational costs associated with slices unlikely to be accepted by 
considering a sphere inscribed in the polyhedron and concentric with its 
center of mass, for which circular regions exposed by a fracture plane bound the area of a prospective face  
from below, with the corresponding acceptance probability overestimating that of the prospective new facet.  In this manner, we 
avoid consideration of fracture planes for which the acceptance probability is no greater than $10^{-8}$, events with negligible 
incidence over the course of an erosion sequence.  Operating in this manner, the computational burden of a sustained slice scales 
no more rapidly than the number of facets of the solids we consider.

The areas of individual facets and the total volume are key quantities in finding the probability of 
a prospective fracture event as well as many of the observables we calculate.  
For individual faces,
the arithmetic mean of each of the member vertex coordinates provides a convenient interior point for 
subdividing the polygonal facet into triangles, whose areas are calculated and summed to yield the combined face area.
On the other hand, to find the total volume of the stone one operates in an analogous fashion, partitioning 
the polyhedron into constituent tetrahedra whose combined volume is the shape total volume. 
To this end, we again subdivide faces into triangles; the three vertices of the latter along with an 
interior point (given by the mean of all of the polyhedron member vertices) define the component tetrahedron. 

Complementary to the sphericity measures and the IPR are quantities which bear more specifically on the shape of the stone, namely 
parameters which characterize the degree to which a stone is oblate or prolate.  Measures of the latter are 
gleaned from the principle moments of inertia of the equivalent ellipsoid, obtained by diagonalizing the moment of 
inertia tensor of the stone.  As in the calculation of total volume, the polyhedron is partitioned into constituent 
tetrahedra with elements $I_{ij}$ of the  moment of inertia tensor given by~\cite{Tonon}  
\begin{equation}
I_{ij} = \frac{V_{\mathrm{tet}}}{20} \left [\delta_{ij} \sum_{k=1}^{3}(s_{k}^{2} + t_{kk}) - s_{i}s_{j} - t_{ij} \right ]
\label{eq:Eq137}
\end{equation}
where $V_{\mathrm{tet}}$ is the volume of the constituent tetrahedron,
while $s_{i} \equiv \sum_{l=1}^{4} x_{li}$ and $t_{ij} \equiv \sum_{l=1}^{4} x_{li} x_{lj}$ with, e.g.,  $x_{li}$
being the $i$th component of the $l$th vertex of the tetrahedron.
Combining moments for each component tetrahedron yields the moment of inertia tensor for the shape as a whole.  
To find the moment of inertia relative to the center of mass, as  
appropriate for a \textit{bona fide} ellipsoid, we appeal to the Steiner parallel axis theorem~\cite{Marion} while taking advantage of the 
fact that the center of mass for a tetrahedron of uniform composition is the arithmetic mean of its four vertices
to find the center of mass of the entire shape.
In terms of semi-major axes $\{ a, b, c \}$ of the equivalent ellipsoid 
(i.e. with $a > b > c$), the oblateness and prolateness measures $\psi_{\mathrm{O}}$ and 
$\psi_{\mathrm{P}}$ are taken to be $\psi_{\mathrm{O}} = (b - c)/a$ and
$\psi_{\mathrm{P}} = (a - b)/a$, respectively.

\subsection{Regular and Irregular Protoclasts}

In selecting initial shapes, our simulation program bifurcates, including both regular and irregular 
polyhedra as primordial forms.  In the case of the former, we examine cubes, and the benefit of 
beginning with regular shapes is two-fold, with one useful feature being   
that equilibration in steady state scenarios is more 
rapid for regular polyhedra than for irregular counterparts.  In addition, we find in the case 
of cube shaped protoclasts two structural phase transitions,  in the first of which all remnants of 
the primordial facets are cleaved away simultaneously.  Subsequently, an additional second order transition 
signals a shift to a spherical shape. 

\begin{figure}
\includegraphics[width=.45\textwidth]{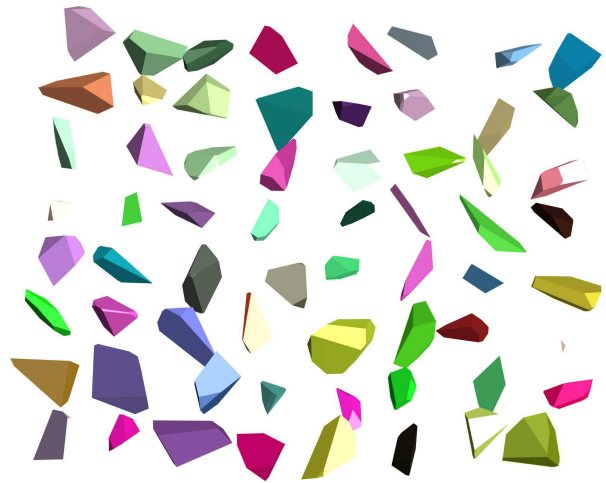}
\caption{\label{fig:Fig1} (Color online) Sixty four sample irregular protoclasts}
\end{figure}

\begin{figure}
\includegraphics[width=.45\textwidth]{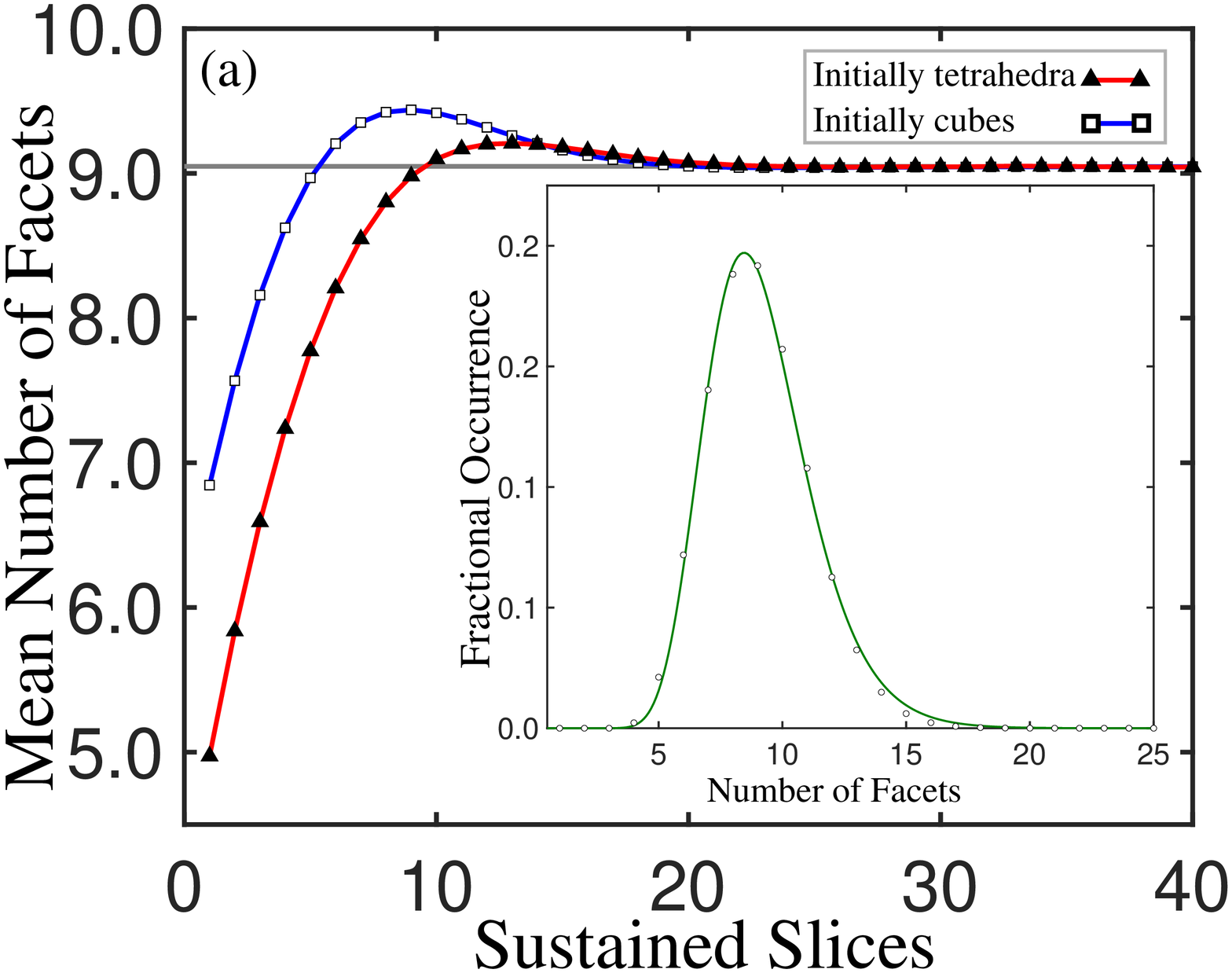}
\includegraphics[width=.45\textwidth]{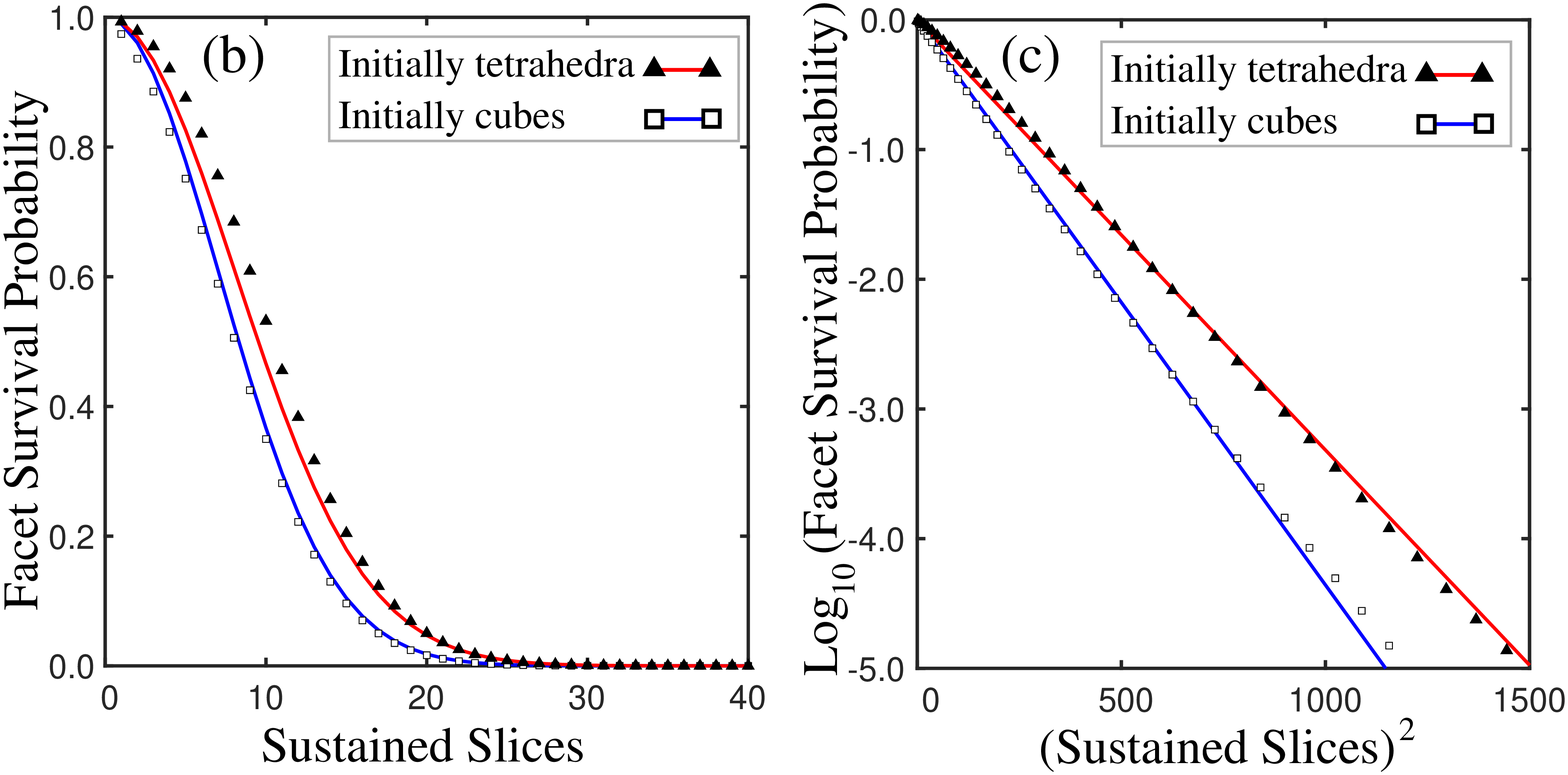}
\caption{\label{fig:Fig2} (Color online) Mean number of facets versus 
sustained slices versus sustained slices for cubic and 
tetrahedral protoclasts (square and triangular symbols respectively), 
with solid curves a guide to the eye. Shown in the inset is the 
frequency facet number distribution; circles are Monte Carlo results while the 
solid line is an optimized fit to a log-normal distribution. Panels (b) and (c)
show the fraction of original facets remaining with solid lines a fit to 
a Gaussian decay in $N_{\mathrm{sust}}$.}
\end{figure}

Cohorts of initially irregular  shapes, on the other hand, are envisaged as a closer geometric match to protoclasts
formed through dramatic fragmentation events.  To generate irregular polyhedra in this very aggressive manner, a cube 
is subject to a sequence of slices, with each prospective slice accepted independent of the area 
$\Delta A$ of the new face. 
Images of sample polyhedra generated in this fashion are shown in Fig.~\ref{fig:Fig1}.
As illustrated in panel (a) of Fig.~\ref{fig:Fig2}, with on the order of 15 sustained slices, 
the mean number of sides quickly converges to 9.03(1) facets on average.
The traces in the main graph, plotted with respect to $N_{\mathrm{sust}}$, were obtained with 
cubes and tetrahedra as initial forms.  The frequency 
distribution of sides, shown  in the graph inset, converges with similar rapidity; in the inset graph,
symbols indicating Monte Carlo results and the solid line being a fit to a log-normal   
distribution with $\sigma = 0.233$ and $\mu = 2.189$.  
We also exhibit in panels (b) and (c) of Fig.~\ref{fig:Fig2} the survival probability $f_{\mathrm{sur}}$ or mean 
likelihood of the persistence of a facet or portion of a facet original to the parent solid.
As indicated in panel (b) of Fig.~\ref{fig:Fig2}, $f_{\mathrm{sur}}$ is strongly suppressed after 
25 sustained slices.  As the semilogarithmic plot with respect to the square of the number of sustained slices in 
panel (c) of Fig.~\ref{fig:Fig2} suggests, the $f_{\mathrm{sur}}$ decay at an approximately Gaussian rate with 
$f_{\mathrm{sur}} \approx e^{-(N_{\mathrm{sust}}/N_{0})^{2}}$ where $N_{0} = 9.98$ for cube shaped initial forms and 
$N_{0} = 11.4$ for tetrahedron shaped protoclasts.
In simulations involving irregular protoclasts, 100 sustained slices are imposed to remove 
any vestige of the seed form.

\section{Steady State Scenario}

\begin{figure}
\includegraphics[width=.45\textwidth]{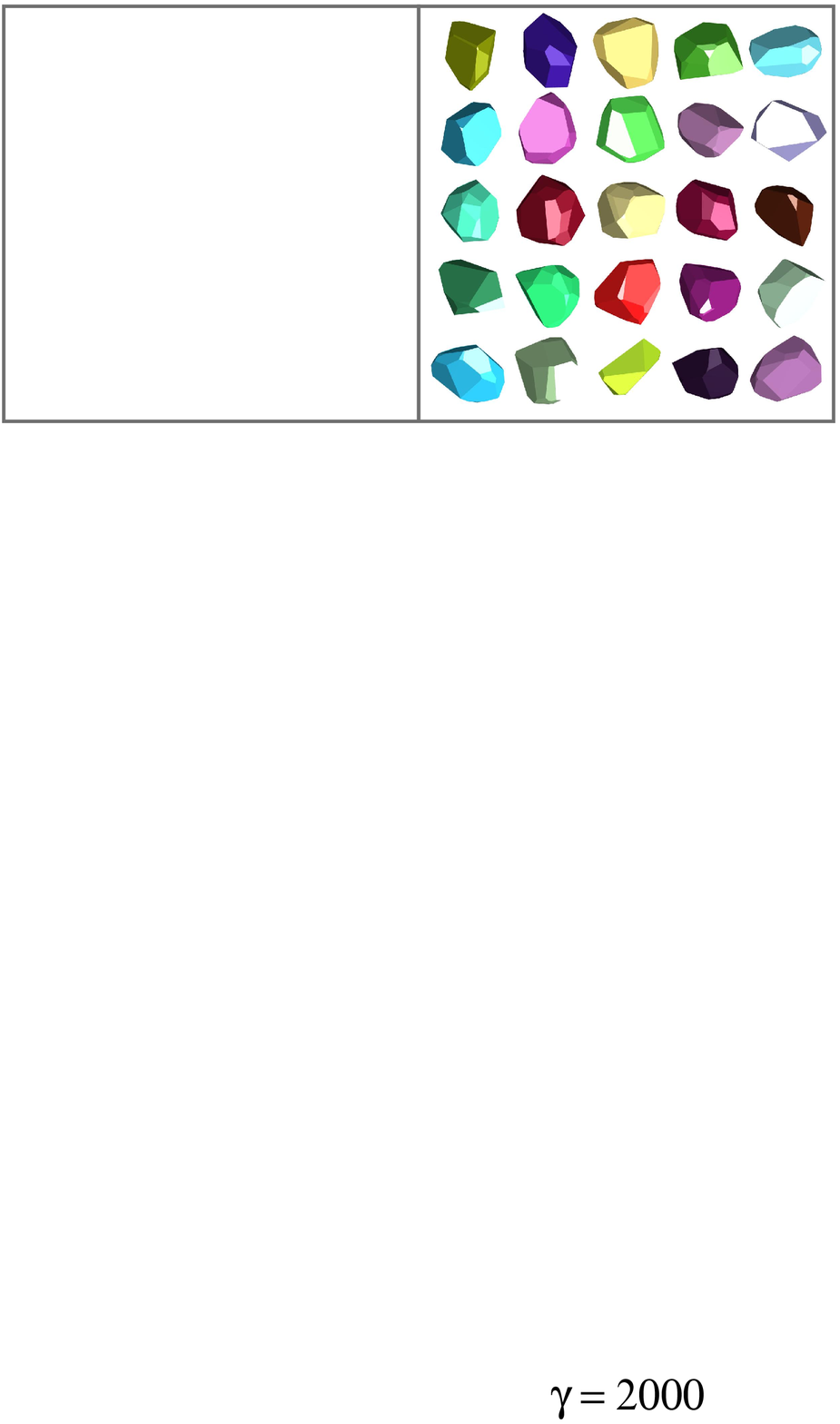}
\caption{\label{fig:Fig3} (Color online) Steady state shapes for various $\gamma$ values}
\end{figure}

For the $\alpha = \alpha_{c}$ case where the acceptance probability for a prospective slice is 
$e^{-\gamma \Delta A/A_{s}}$ ($A_{s} = [6 \pi^{\frac{1}{2}} V]^{\frac{2}{3}}$ being the area of the volume equivalent sphere), 
a steady state may in principle develop with the mean number of 
faces $\langle n \rangle$ converging to a fixed value since $\langle \Delta A \rangle \sim A_{\Sigma}/\gamma$ per the 
acceptance probability relation.  
Images of stones sampled from the steady state ensemble appear in Fig.~\ref{fig:Fig3}, showing
the emergence of smooth spherical shapes for toughness parameters ranging from $\gamma = 10$ to $\gamma = 2000$.

A salient question related to the 
attainment of equilibrium is how long is needed (in terms of the number of 
sustained slices $N_{\mathrm{sust}}$) for an ensemble of stones to reach steady state.  
In the context of numerical simulations, we insist that all observables of interest remain constant with respect 
to doubling of $N_{\mathrm{sust}}$.  For $\gamma = 100$, 3,000 sustained slices satisfies this 
criterion.  As we now argue from geometric considerations,
for $\gamma \gg 1$ (i.e. certainly for $\gamma \geq 100$) 
time scales for the removal of equivalent fractions of the original volume of an ensemble of 
rocks (and hence salient stages such as the achievement of steady state) scale quadratically with 
the toughness parameter $\gamma$.  

Although the acceptance probability sets the scale $\langle \Delta A \rangle$ for the typical area of a new facet,
fractures leaving behind faces with the same area need not cleave 
away the same volume.  One anticipates more volume to be removed from sharply peaked features  than
from regions with less curvature by the same number of sustained slices~\cite{Jerolmack1}.
Thus the edges and vertices of the 
primordial polyhedron, 
are quickly worn down in the initial stages of the erosion process.

In the $\gamma \gg 1$ regime, fracture planes are constrained to be shallow,
unable to penetrate deeply due to the exponential penalty on $\Delta A$ in the 
acceptance probability.  We introduce the cleaving plane, just below the rock surface, 
as well as the parallel plane of tangency with a single point of contact with the surface,  
while superimposing Cartesian axes with the $xy$ plane coinciding with the tangent plane and the $z$ axis  
directed toward the interior of the stone.  Hence, for the rock surface one would generically have
have $f(x,y) \approx A x^{2} + 2 B xy + C y^{2}$ near its minimum at the point of tangency, where $A$, $B$, and $C$ 
are locally determined constants.  Seeking to eliminate the diagonal term, we have in matrix form
\begin{equation}
f(x,y) = \left [ \begin{array}{cc} x & y \end{array} \right ] \left[ \begin{array}{cc} A & B \\ B & C \end{array} \right ]  
\left[ \begin{array}{c} x \\ y \end{array} \right ];~~\vec{v}_{\mathrm{plane}} \equiv \left [ \begin{array}{c} x \\ y \end{array} \right ]
\end{equation}
where $\vec{v}_{\mathrm{plane}}$ is the component in the tangent plane. 
The eigenvalues of the symmetric matrix are:
$\lambda_{1,2} = \frac{1}{2}[(A + C) \pm \sqrt{ (A - C)^{2} + 4 B^{2}} ]$ 
with $\lambda_{1,2} > 0$ if $AC > B^{2}$, as must be the case for the convex objects we consider. In terms of
the orthonormal eigenvectors $\hat{u}_{1}$ and $\hat{u}_{2}$, one may write 
$\vec{v}_{\mathrm{plane}} = \alpha_{1} \hat{u}_{1} + \alpha_{2} \hat{u}_{2}$ and thus
\begin{equation}
f(x,y) \rightarrow g(\alpha_{1},\alpha_{2}) = \lambda_{1} \alpha_{1}^{2} + \lambda_{2} \alpha_{2}^{2}
\end{equation}

The cleaving plane hence exposes an elliptical region for which
$z = \lambda_{1} \alpha_{1}^{2}
+ \lambda_{2} \alpha_{2}^{2}$ for a fixed depth $z$ beneath the point of tangency,  
where the area is $A(z)  = \pi a b = \pi z/l$ 
with $l \equiv (\lambda_{1} \lambda_{2})^{-1/2}$ being a locally defined length scale.   
Thus, the depth of a typical fracture plane (i.e. with an area $\langle A \rangle$) 
is $D = \langle \Delta A \rangle  (\pi l)^{-1}$, and the typical volume $\Delta V$ removed
is $\Delta V = \int_{0}^{D} A(z) dz =  \langle \Delta A \rangle^{2}/(2 \pi l) \sim \gamma^{-2} A_{\mathrm{\Sigma}}/(2 \pi l)$ 
since $\langle \Delta A \rangle \sim A_{\Sigma} \gamma^{-1}$ for the relative area scenario.
Thus, time scales for the removal of volume scale as $\gamma^{2}$ in the steady state scenario and more broadly 
as the square of the toughness parameter in non-steady state schemes as well.

\subsection{The Independent Facet Model}
 
In the steady state calculations, we obtain configuration averaged observables (i.e. for at least 1000 distinct 
realizations) for two and a half decades 
of the mean number $\langle n \rangle$ of facets, finding power law decays in $\gamma$ and 
$\langle n \rangle$ for $\gamma \gg 1$ for 
all measures of the deviation from perfect spherical shapes.  
As a model for quantities of interest, we assume quasi-spherical shapes for cohorts of stones in 
steady state with properties (e.g. areas) of polyhedron faces assumed to be statistically uncorrelated;
in this Independent Facet Model (IFM), characteristics of polyhedron faces are considered to 
statistically uncorrelated; nevertheless, we find good quantitative agreement 
with salient observables from Monte Carlo simulations.

For a spherical geometry, the relative area of a fresh facet is
$\Delta A/A = (2 \tilde{u} - \tilde{u}^{2})/4$ where $\tilde{u} = u/R$, with $R$ being the mean radius of 
the stone and $u$ the distance of the planar slice below the surface of the rock. 
With the probabilities of new faces being exponentially suppressed in $\Delta A$, 
mean facet variables $\langle f(\tilde{v}) \rangle$ are given by
\begin{equation} 
\langle f( \tilde{u} ) \rangle = 
\frac{\int_{0}^{1} f(\tilde{u}) e^{-\gamma(2 \tilde{u} - \tilde{u}^{2})/4} d \tilde{u}}{\int_{0}^{1} e^{-\gamma (2 \tilde{u} - 
\tilde{u}^{2})/4} d \tilde{u}}
\label{eq:Eq100}
\end{equation}
where the denominator plays a role analogous to that of the partition function in statistical mechanics; for 
the sake of obtaining closed form relations accurate to leading or next to leading order in $\gamma$ in for
$\gamma \gg 1$,  Eq.~\ref{eq:Eq100} reduces to 
\begin{equation}
\langle f(\tilde{u}) \rangle \approx \frac{1}{2} \gamma \int_{0}^{\infty} f(\tilde{u}) e^{-\gamma \tilde{u}/2} d \tilde{u} 
\label{eq:Eq150}
\end{equation}
\begin{figure}
\includegraphics[width=.45\textwidth]{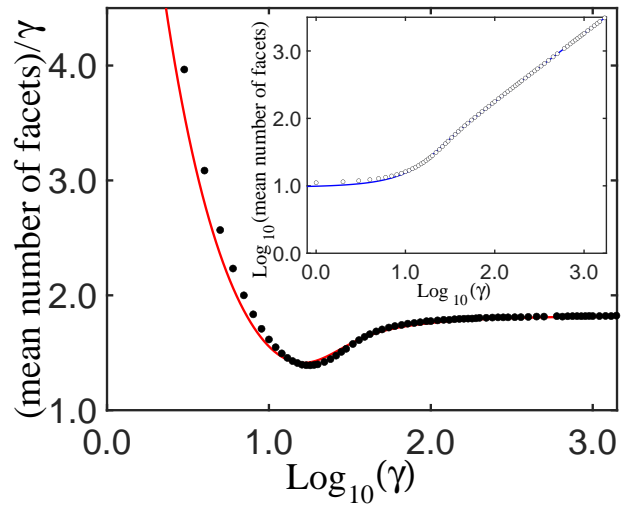}
\caption{\label{fig:Fig4} (Color online) Mean number of facets with respect to $\gamma$. Symbols 
are Monte Carlo results, while the solid line in the main graph and inset is an analytical curve.}
\end{figure}

Fig.~\ref{fig:Fig4} shows the mean facet number $\langle n \rangle$ with  respect to $\gamma$, with symbols representing 
Monte Carlo results; analytical results indicated by the solid curve are in good agreement with the latter.
The main graph is a plot of the ratio $\langle n \rangle/\gamma$ with respect to $\log_{10}(\gamma)$, which converges to 
1.82(1) in the large $\gamma$ regime, while the inset graph is a log-log plot of  $\langle n \rangle$ versus
$\gamma$.  A salient feature is the comparatively gradual increase in $\langle n \rangle$ for $\gamma \leq 10$ with 
a more rapid asymptotically linear rise thereafter.    

The analytical curve is obtained assuming a truncation over time of newly exposed faces, where in terms of the 
mean area $\langle A_{0} \rangle$ of  newly created faces, the average area of $\langle A \rangle$ over its 
lifetime is $\langle A \rangle = \eta \langle A_{0} \rangle$ where $\eta < 1$.
The mean solid angle subtended by a facet, modeled as a circular shape  (with the reduction in area taken into account) is
$f_{\Omega}(\tilde{u}) = 2 \pi [1 - \sqrt{1 - \eta(2 \tilde{u} - \tilde{u}^{2})}]$
where the mean number of facets is hence $\langle n \rangle = 4 \pi/\langle f_{\Omega} (\tilde{u} ) \rangle$.  

One fixes $\eta$ by appealing to the large $\gamma$ limit where $\tilde{u} \ll 1$ due to the 
shallowness of typical sustained slices.   
In this regime, one may expand the radical in $f_{\Omega}(\tilde{u})$ expression; using Equation~\ref{eq:Eq150}, we 
find $\langle n \rangle \approx \gamma/\eta$,  $\eta = 0.549(3)$ and  
$\langle n \rangle = 1.82(1) \gamma$ for $\gamma \gg 1$.

\begin{figure}
\includegraphics[width=.45\textwidth]{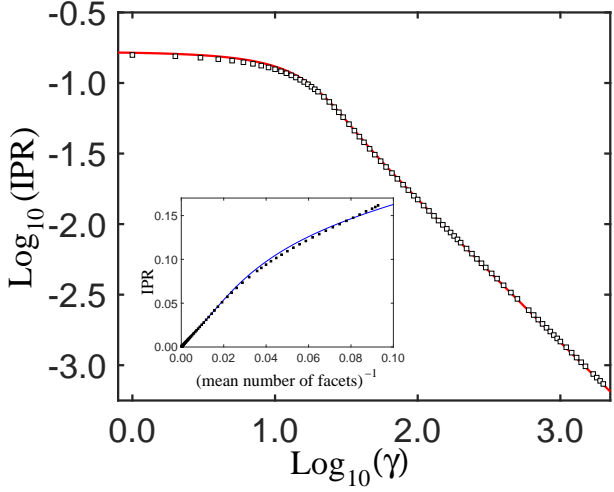}
\caption{\label{fig:Fig5} (Color online) Inverse Participation Ratio (IPR) log-log
plot (main graph) and IPR with respect to the reciprocal of the mean facet number 
(inset).  Symbols represent Monte Carlo data, while the solid curve is from an
analytical model.}
\end{figure}

The Inverse Participation Ratio (IPR) is plotted in Fig.~\ref{fig:Fig5} with the solid curve obtained in the Independent Facet Model framework 
and symbols representing Monte Carlo results in both the inset and the main graph with good quantitative 
agreement among the latter and the former.
In the case of the inset, the IPR is plotted with respect to $\langle n \rangle^{-1}$, suggesting an extrapolation to zero as the number of 
facets becomes infinitely large and thus smooth surfaces in the large $\gamma$ limit.  
On the other hand, the linear decrease of IPR in the log-log curve for $\gamma \geq 10$ in the main graph is compatible 
with an asymptotically power law decrease (specifically as $\gamma^{-1}$ or $\langle n \rangle^{-1}$)  for the Inverse Participation Ratio.

\begin{figure}
\includegraphics[width=.45\textwidth]{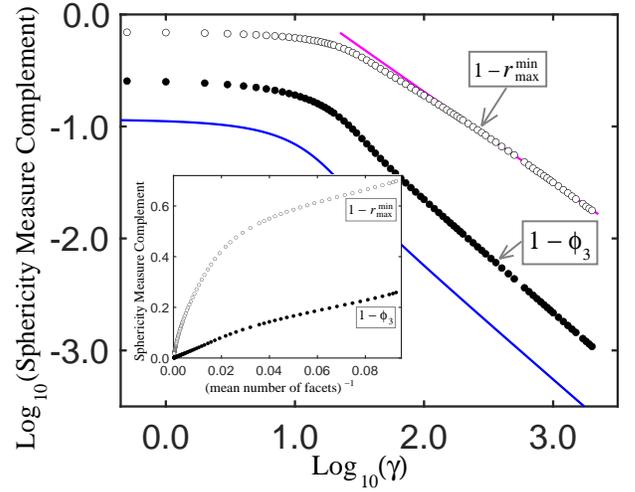}
\caption{\label{fig:Fig6} (Color online) Spherical measure complements $1 - r^{\textrm{min}}_{\textrm{max}}$ (open circles)
$1 - \phi_{3}$ (filled circles) log log 
plot (main graph) and results plotted with respect to the reciprocal of the mean facet number
(inset). The solid blue curve in the main graph is an analytical curve from the 
Independent Facet Model, while the magenta line fitted to the $1 - r_{\mathrm{max}}^{\mathrm{min}}$ is a power law decay scaling as 
$\gamma^{-\zeta}$ where $\zeta = 0.81$.}
\end{figure}

In the IFM framework, we posit that $\langle \sum_{i = 1}^{n} A_{i} \rangle$ may be replaced with $\langle A \rangle \langle n \rangle$ and 
$\langle \sum_{i = 1}^{n} A_{i}^{2} \rangle$ with $\langle A^{2} \rangle \langle n \rangle$, yielding
\begin{equation}
\textrm{IPR} \approx \frac{\chi \langle f(2 \tilde{u} - \tilde{u}^{2}) \rangle }{ \langle n \rangle \langle f([2 \tilde{u} - \tilde{u}^{2}]^{2}) \rangle} 
\end{equation}
where $\chi = 1.32(1)$ is a dimensionless fitting parameter on the order of unity. The analytical IPR curve obtained in this manner 
with the mild rescaling by $\chi$ is in excellent quantitative agreement with Monte Carlo simulation results as may be seen in Fig.~\ref{fig:Fig6}.
By appealing to Equation~\ref{eq:Eq150} in the $\gamma \gg 1$ regime, we find that 
the inverse participation ratio tends to:  $\textrm{IPR} = (2 \chi \eta) \gamma^{-1} =  (2 \chi) \langle n \rangle^{-1}$.

Sphericity complement measures are shown in Fig.~\ref{fig:Fig6} for $1 - \phi_{3}$ (filled circles) and $1 - r^{\mathrm{min}}_{\mathrm{max}}$
(open circles) on a log-log scale with respect to $\gamma$ in the main graph and versus $\langle n \rangle^{-1}$ in the inset graph.  From 
the main graph, one see that both complements shift to power law decays after a plateau qualitatively similar to that other IPR in 
Fig.~\ref{fig:Fig5}, with $1 - \phi_{3}$ decreasing more rapidly than $1 - r^{\mathrm{min}}_{\mathrm{max}}$ with the asymptotic downward 
slope in the log-log graph appearing to be greater for the former than for the latter.
The trend to zero of departures from a perfect spherical shape with increasing number of facets is also evident in the graph inset 
which shows $1 - \phi_{3}$ and $1 - r_{\mathrm{max}}^{\mathrm{min}}$ 
with respect to the reciprocal of the mean number of facets, $\langle n \rangle^{-1}$.

In the IFM framework, we obtain the correct asymptotic behavior in the case of the sphericity complement 
$1 - \phi_{3}$.  In this vein, one begins by considering a sphere with a volume of $\frac{4}{3} \pi R^{3}$, and $\langle n \rangle$ lenticular 
slices are sheered away by the fracture planes;  one calculates in the IFM context 
the mean of  the ratio of the surface area of the volume equivalent sphere to the area of this 
solid,  namely $\langle [(2/s)(1 - s)(1 - \sqrt{1 - s})]^{\frac{1}{3}} \rangle$ where $s \equiv \eta (2 \tilde{u} - \tilde{u}^{2})$.
The result is the solid curve in the main graph of~\ref{fig:Fig6}.
Although depressed relative to the $1 - \phi_{3}$ curve, the IFM result and Monte Carlo simulation results nonetheless have in common 
a plateau-like region for $\gamma \leq 10$ which gives way to a linear slope signaling a power law decay.   In fact, from Eq.~\ref{eq:Eq150}, 
the steady state sphericity complement $1 - \phi_{3}$ scales as $\frac{4}{3} \eta \gamma^{-1}$,
asymptotically correct apart from a dimensionless prefactor.

Whereas $1 - r_{\mathrm{max}}^{\mathrm{min}}$ is likely sensitive to correlations among facet sizes and positions, and thus less amenable to
the IFM treatment, a statistical analysis of the time dependence of the $1 - r_{\mathrm{max}}^{\mathrm{min}}$ suggests that it saturates 
at values proportional to $\gamma^{-\zeta}$ where $\zeta = 0.81(5)$.
This power law decay is indicated by the magenta line in the main graph of Fig.~\ref{fig:Fig6}.

\begin{figure}
\includegraphics[width=.45\textwidth]{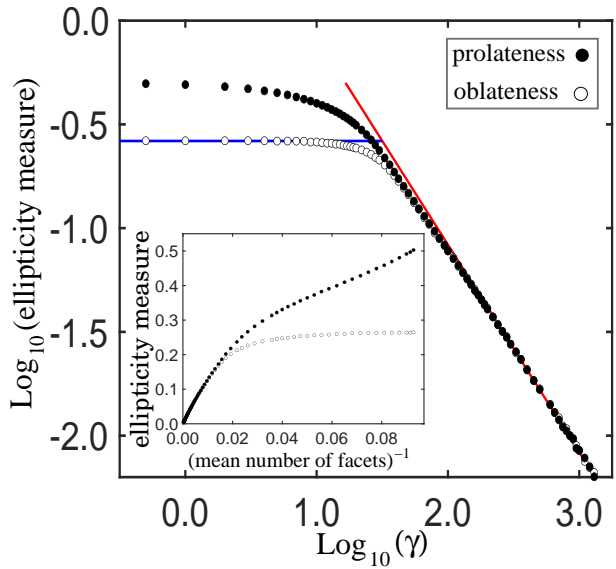}
\caption{\label{fig:Fig7} (Color online) Log-log plot of ellipticity measures, including
oblateness and prolateness with respect to 
the mean facet number $\langle n \rangle^{-1}$  Simulation results are indicated with symbols, while the 
red line on the right side corresponds to a power law scaling of $\gamma^{-1}$. In the inset, the 
ellipticity measures are plotted with respect to $\langle n \rangle^{-1}$.}
\end{figure}

Measures of ellipticity, the oblateness (open symbols) and prolateness (filled symbols), 
are shown in Fig.~\ref{fig:Fig7} in a log-log plot. Both the oblateness and the prolateness decrease gradually with $\gamma$, 
for $\gamma \leq 10$, with the former initially flat as indicated by the blue line.  Asymptotically linear on a log-log scale, the prolateness and 
oblateness converge in the large $\gamma$ regime. The red line highlights the power law decay of 
both ellipticity measures, corresponding to a $1/\gamma$ dependence.

\subsection{Time Evolution of Shapes}

\begin{figure}
\includegraphics[width=.45\textwidth]{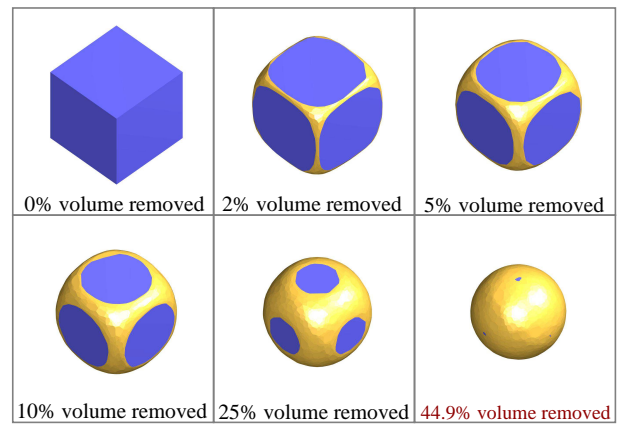}
\caption{\label{fig:Fig8} (Color online) Images of stones derived from
a cube shaped protoclast at various stages of 
erosion for $\gamma = 2000$, with facets original to the parent cube 
shown in blue.  The structural transition appears at the extreme lower right
corner.} 
\end{figure}

\subsubsection{Structural Phase Transitions in Mono-dispersed Protoclasts}
Although steady state shapes are smooth and spherical for $\gamma \gg 1$, the initial 
cohort of protoclasts are polyhedral with a relatively small number of facets.  We consider regular cubes as 
protoclasts as well as the highly irregular initial shapes mentioned earlier.  In the case 
of the former, we observe critical behavior as primordial facets are eroded away, and 
stones become smoother and rounder as additional volume is chipped away; this transition 
from shapes which possess facets form the parent solid and those which do not is abrupt with all 
of the hallmarks of a second order phase transition and is reflected in all of the variables we  
calculate apart from measures of deviation from a spherical shape.  The latter exhibit critical behavior
at a subsequent phase transition in which the stones revert to spherical shapes.  

Selected images from the erosion trajectory of cube shaped protoclasts appear in
Fig.~\ref{fig:Fig8} with the image at the lower right coinciding with the structural transition in which
all primordial facets are removed; blue areas are facets or portions of facets original to the parent solid.
On the other hand, due to the inherent strong structural disorder of the irregular protoclasts, 
the disappearance of primordial facets occurs at different times for different 
shapes, with a concomitant loss of any well defined phase transition for ensemble averaged
observables for the cohort as a whole.
In fact, by considering mono-dispersed cohorts made up of a single irregular shape, 
we see that the elimination of protoclast faces is in general asynchronous even for 
an ensemble of initially identical shapes, with a simultaneous elimination of primordial facets 
limited to highly symmetric shapes and being the 
exception rather than the rule.

\begin{figure}
\includegraphics[width=.45\textwidth]{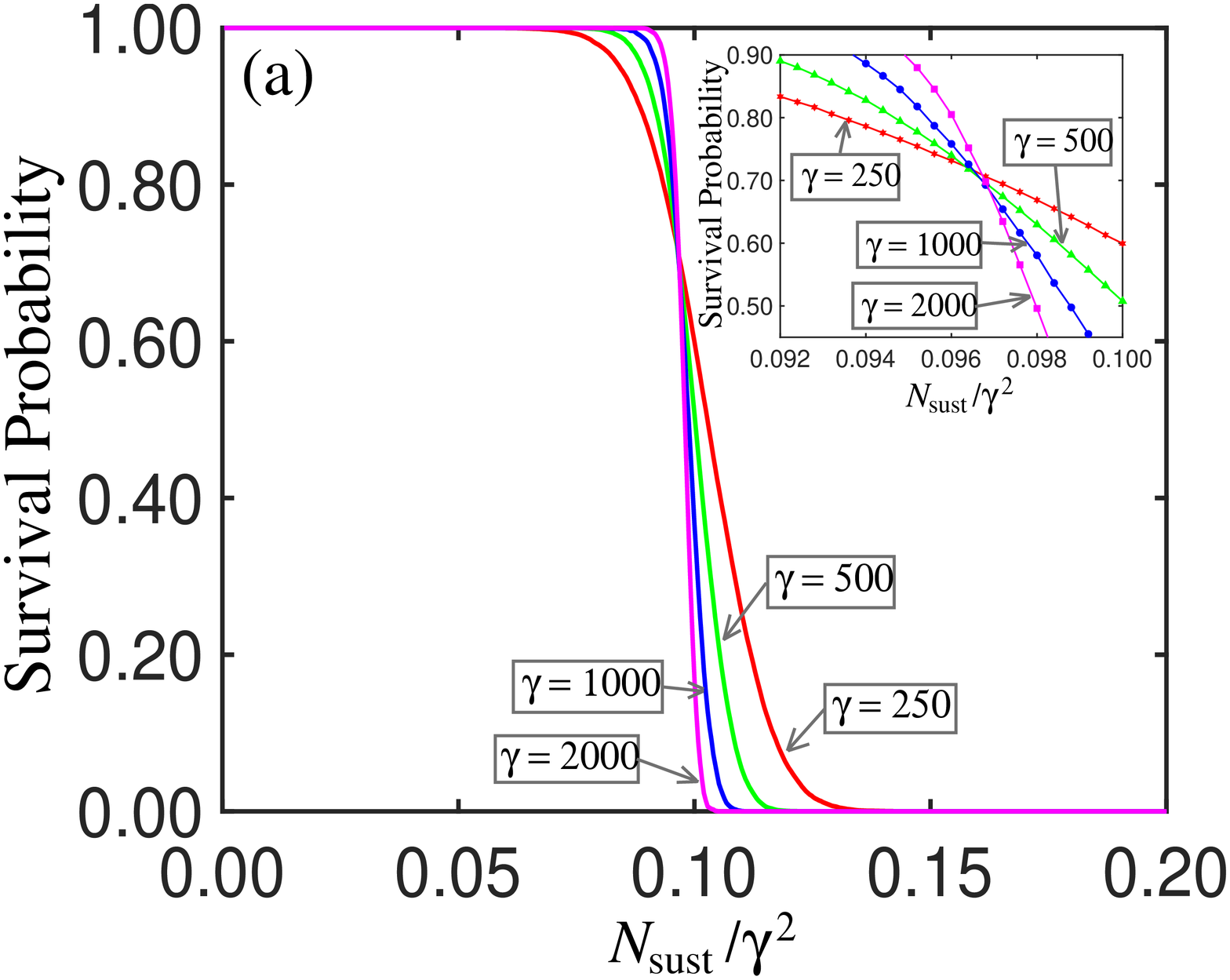}
\includegraphics[width=.45\textwidth]{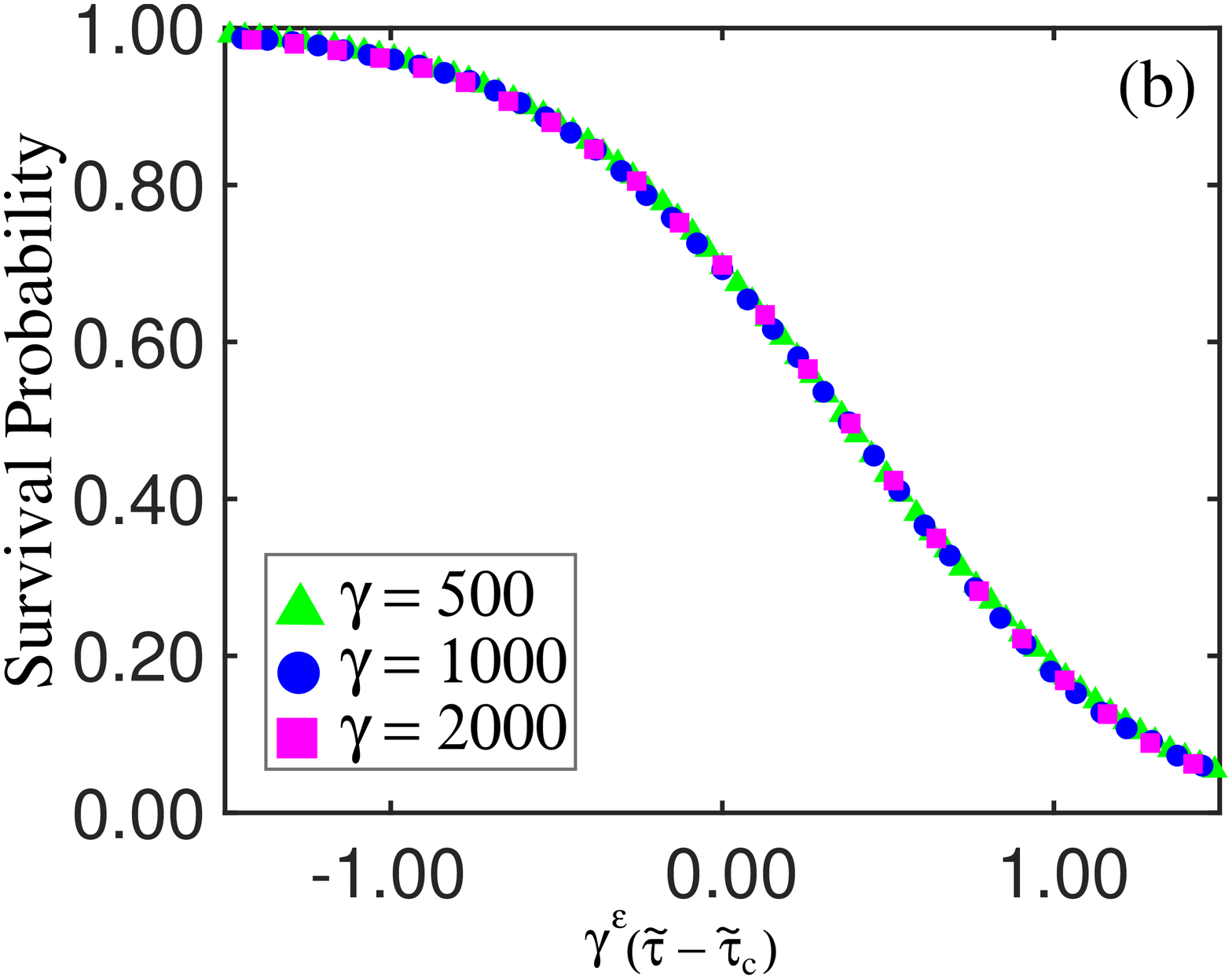}
\caption{\label{fig:Fig9} (Color online) Facet survival probabilities $f_{\mathrm{sur}}$ for cube 
shaped protoclasts are shown in panel (a) with 
respect to $\tilde{\tau} = N_{\mathrm{sust}}/\gamma^{2}$;a closer view of the 
$f_{\mathrm{sur}}$ curves in the inset.  Panel (b) shows an optimized 
data collapse for cube shaped protoclasts with $\varepsilon = 0.76(5)$.}
\end{figure}

In statistical mechanical analyses of phase transitions, a standard 
practice is to specify an order parameter, a variable which is finite when 
the phase in question is present, and zero otherwise if one is in the bulk or 
thermodynamic limit.  For this purpose, we use the survival probability (denoted in this 
work as $f_{\mathrm{sur}}$), 
or the ensemble averaged fraction of primordial facets remaining.  

The survival probability for cube shaped protoclasts is shown in in Fig.~\ref{fig:Fig9} for 
$\gamma$ values ranging from $\gamma = 250$ to $\gamma = 2000$. With characteristic times 
scaling as $\gamma^{2}$, we plot the survival probability and other 
salient variables with respect to the scaled time,
$N_{\mathrm{sust}}/\gamma^{2}$, also denoted as $\tilde{\tau}$.  
The mean fraction of surviving original facets initially exhibits a plateau, 
decreasing to zero as all of the primordial cube facets are eroded away.  That  
the descent of the survival index sharpens as $\gamma$ increases 
[with curves crossing at a common point as shown in the inset of panel (a) of Fig.~\ref{fig:Fig9}]
 is compatible with the original facet survival probability being a viable order parameter 
(i.e. non-zero only when vestiges of the original protoclast faces are still present) in the 
context of a second order phase transition at a critical $\tilde{\tau}$ value,
$\tilde{\tau}_{c}$.  

Singularities in observables or their derivatives in the vicinity of $\tilde{\tau}$ 
are hallmarks of a second order phase transition, which we use to show that salient   
variables are influenced by critical behavior, and to calculate indices of 
interest such as $\tilde{\tau}_{c}$.  In the context of our simulations, the thermodynamic limit  
corresponds to $\gamma \gg 1$, where the number of facets also is large, (e.g. as high as 3,600 
for $\gamma = 2000$).

As is customary in the analysis of 
second order phase transition, we elucidate critical behavior in the structural phase 
transition of eroding cubes by appealing to single parameter finite size scaling 
theory, where one uses the data collapse phenomenon as a quantitative tool to calculate 
critical indices.  Although in general one would plot $\gamma^{\beta} f_{\mathrm{sur}}$ with 
respect to $\gamma^{\varepsilon} (\tilde{\tau} - \tilde{\tau}_{c})$, the crossing of the 
$f_{\mathrm{sur}}$ curves for different $\gamma$ values suggests that $f_{\mathrm{sur}}$ is of 
zero scaling dimension, as is true for the Binder Cumulant~\cite{Binder} used in connection with 
thermally driven magnetic phase transitions, for which $\beta = 0$. Accordingly,
$f_{\mathrm{sur}}$ is on the ordinate axis  
in panel (b), with the optimal data collapse occurring for $\tau_{c} = 0.0968(2)$
and $\varepsilon = 0.76(5)$.

\begin{figure}
\includegraphics[width=.45\textwidth]{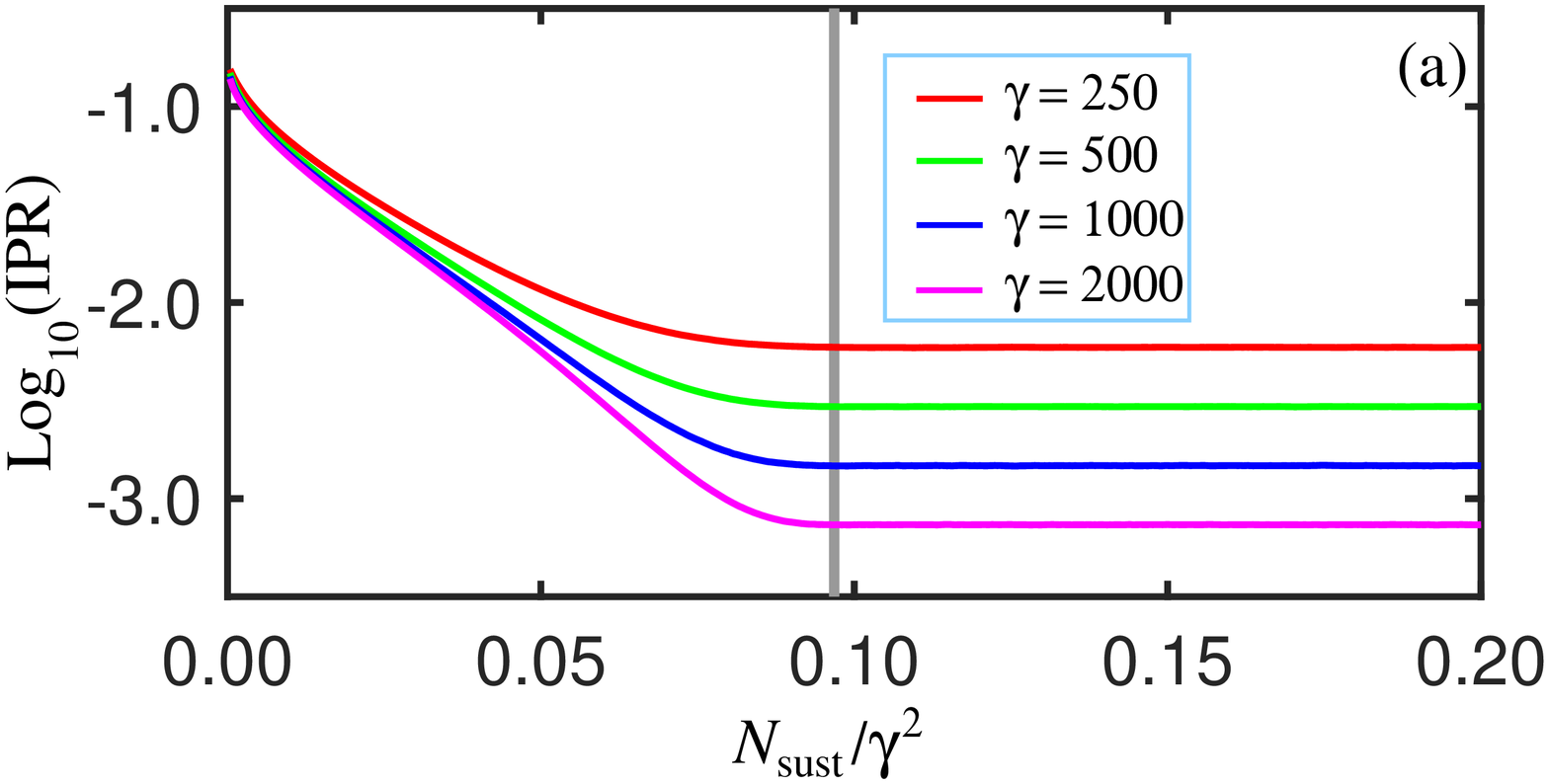}
\includegraphics[width=.45\textwidth]{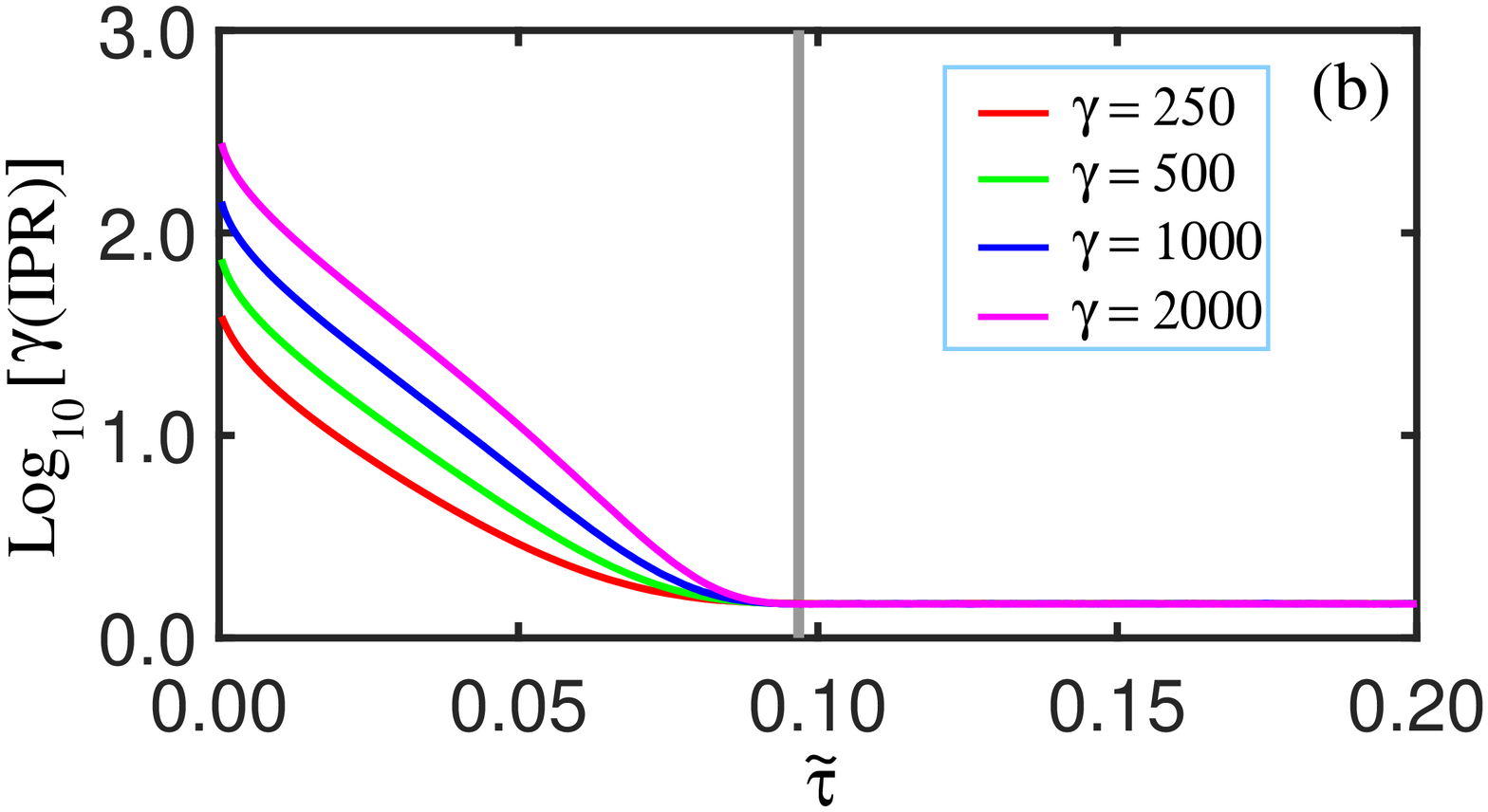}
\includegraphics[width=.45\textwidth]{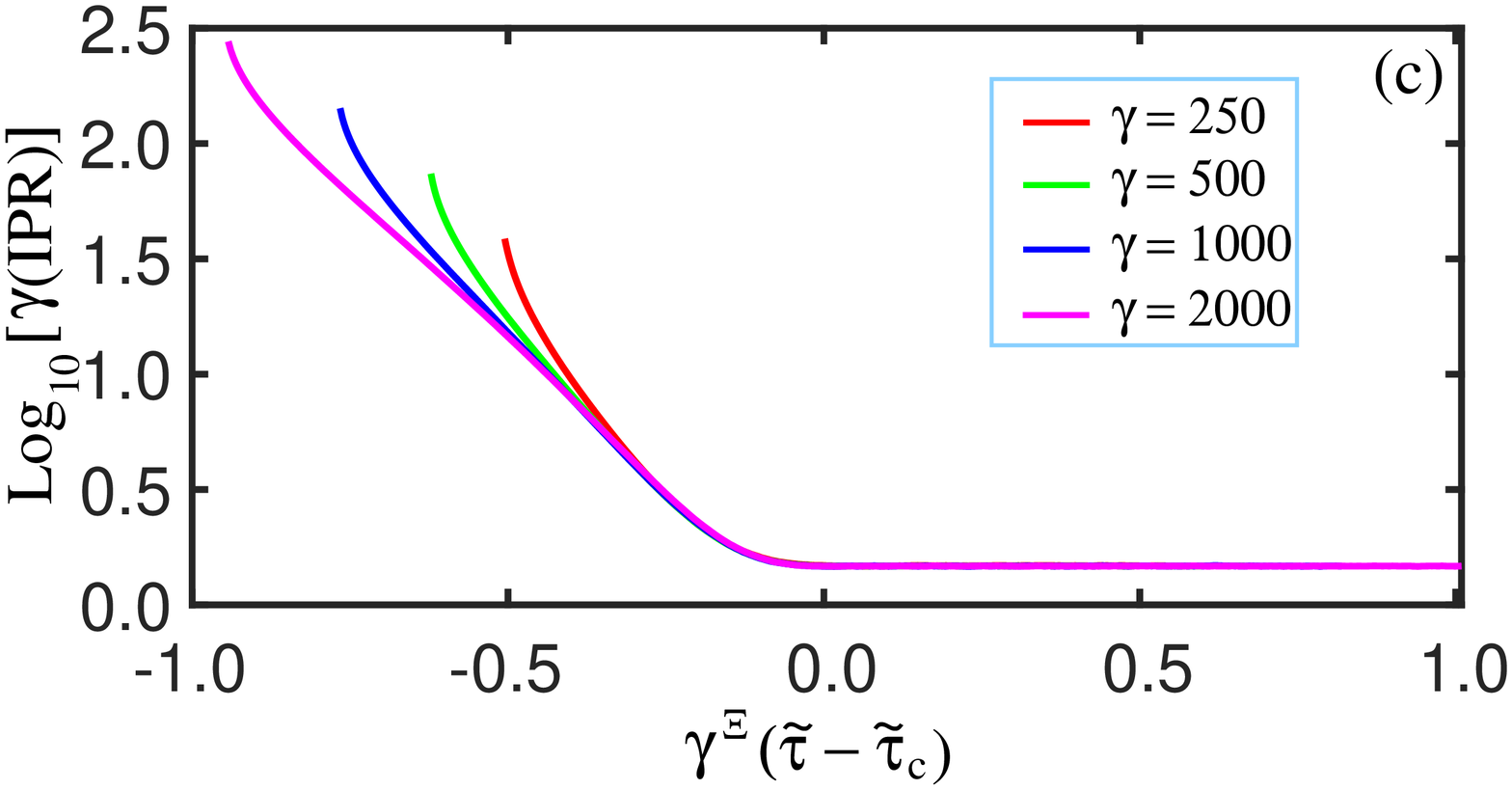}
\caption{\label{fig:Fig10} (Color online) Inverse Participation Ratios (IPR) for cube shaped protoclasts
in panel (a) for various $\gamma$ values.  
Panel (b) is a plot of $\gamma \times \textrm{IPR}$ with respect to $\tilde{\tau}$ on a semi-logarithmic scale 
with a splaying of the curves for $\tilde{\tau} < 0.097$.
Panel (c) is a data collapse plot with $\gamma \times \textrm{IPR} $ 
plotted with respect to $\gamma^{\Xi}(\tilde{\tau} - \tilde{\tau}_{c}$) where $\Xi = 0.30(5)$.}
\end{figure}

The Inverse Participation Ratio (IPR), a measure of the smoothness of stones, is shown in panel (a) of Fig.~\ref{fig:Fig10}
on a semi-logarithmic scale for cube shaped protoclasts.  
The regular spacing on the logarithmic scale of the asymptotically flat IPR curves for large enough $\tilde{\tau}$ is compatible with the 
Participation Ratio scaling as $\gamma^{-1}$ at steady state, also evident in panel (b) of Fig.~\ref{fig:Fig10} where
$\gamma \times \textrm{IPR}$ is plotted with respect to $\tilde{\tau}$.  For $\tilde{\tau} > \tilde{\tau}_{c}$, the curves for various 
$\gamma$ values merge onto a single flat line, splaying outward for $\tilde{\tau} < \tilde{\tau}_{c}$, suggesting singular behavior for $\tilde{\tau} = 0.0967$.
In panel (c) of Fig.~\ref{fig:Fig10}, the quantity $\gamma \times \textrm{IPR}$ is plotted with respect to $\gamma^{\Xi}(\tilde{\tau} - \tilde{\tau}_{c})$ 
where $\Xi = 0.30(5)$ [and $\tilde{\tau} = 0.097(1)$ as in the case of $f_{\mathrm{sur}}$] 
for various $\gamma$ values, with the overlap of the curves indicating a collapse of the entire gamut of the 
IPR onto a universal curve.

\begin{figure}
\includegraphics[width=.45\textwidth]{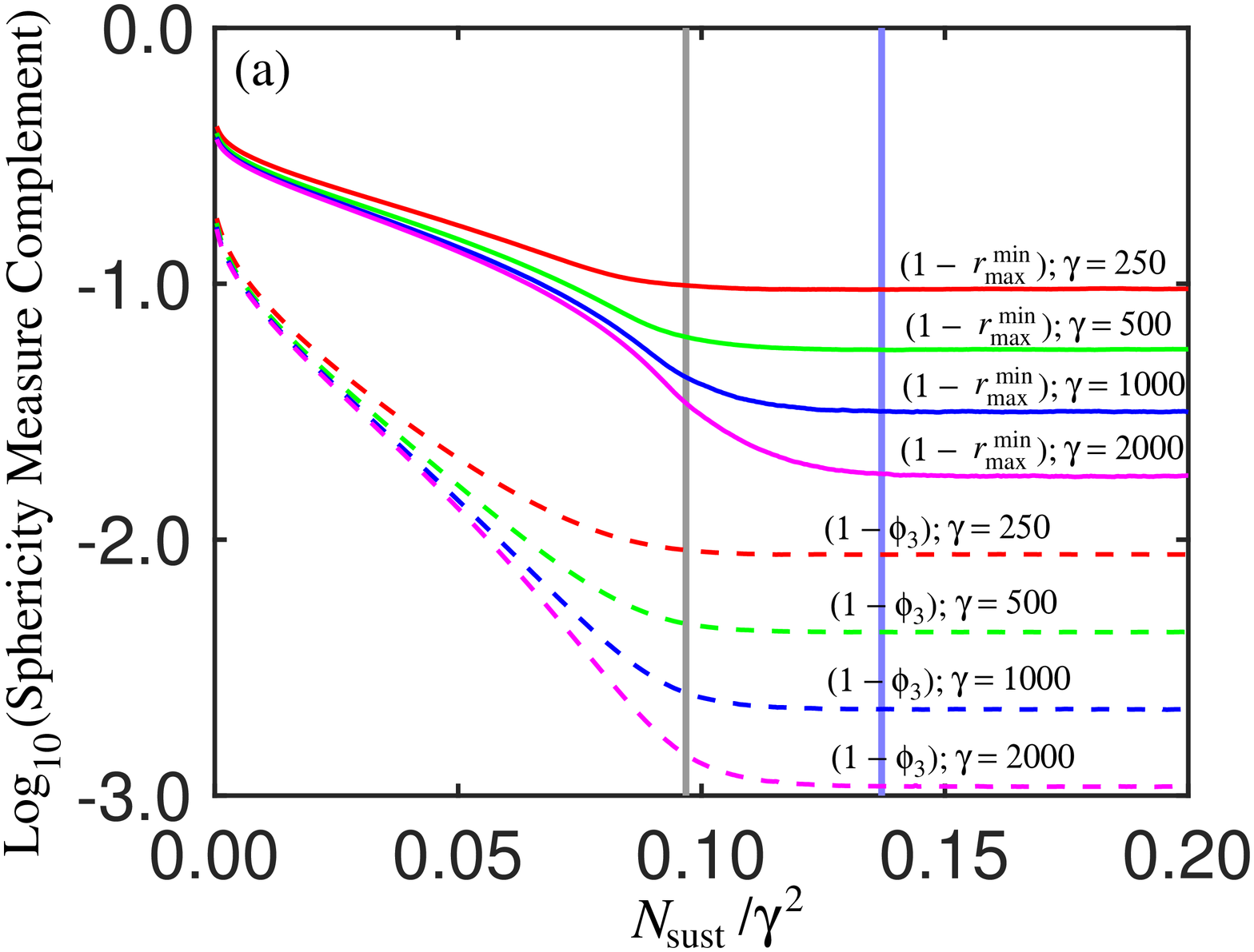}
\includegraphics[width=.45\textwidth]{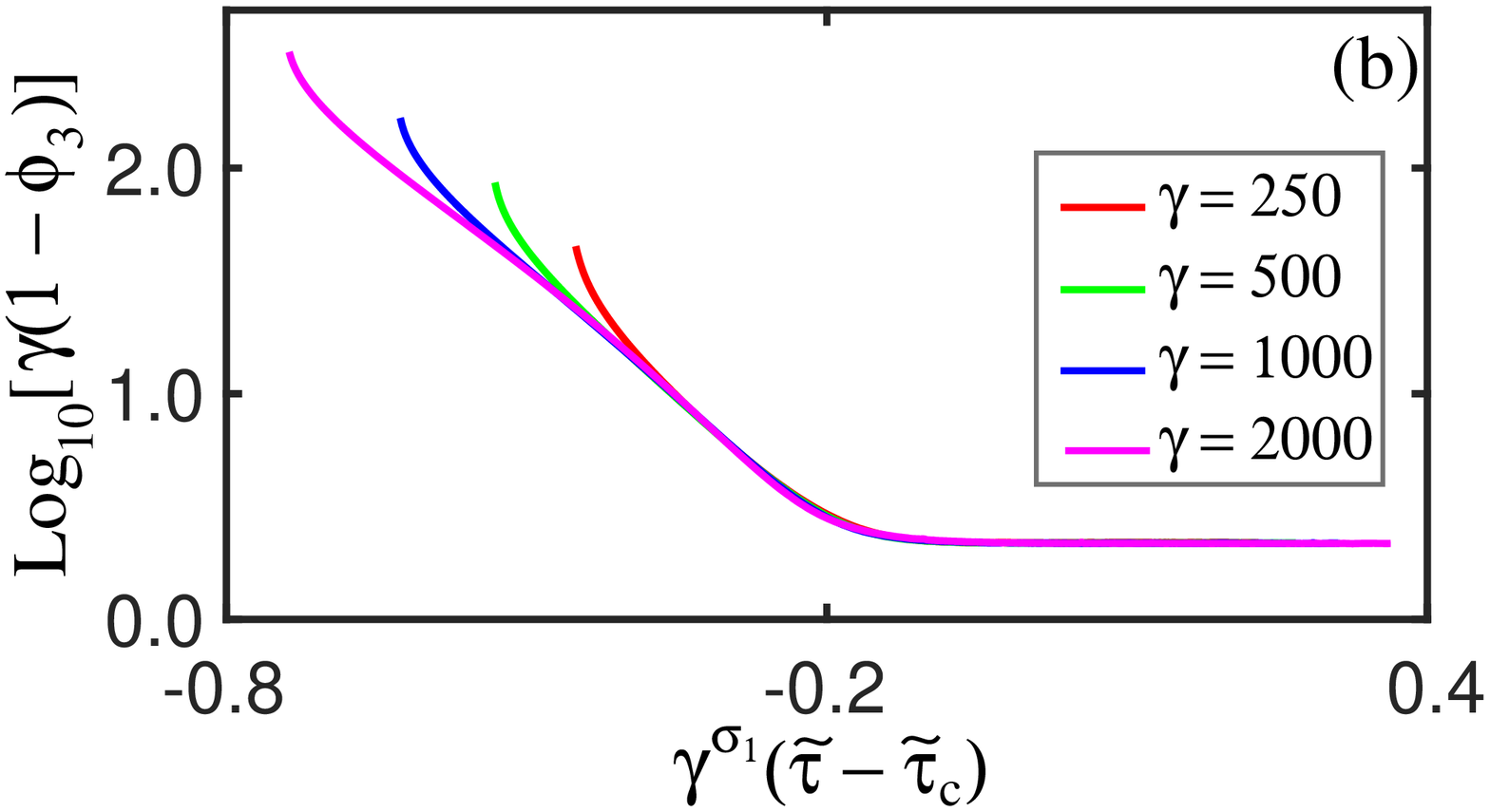}
\includegraphics[width=.45\textwidth]{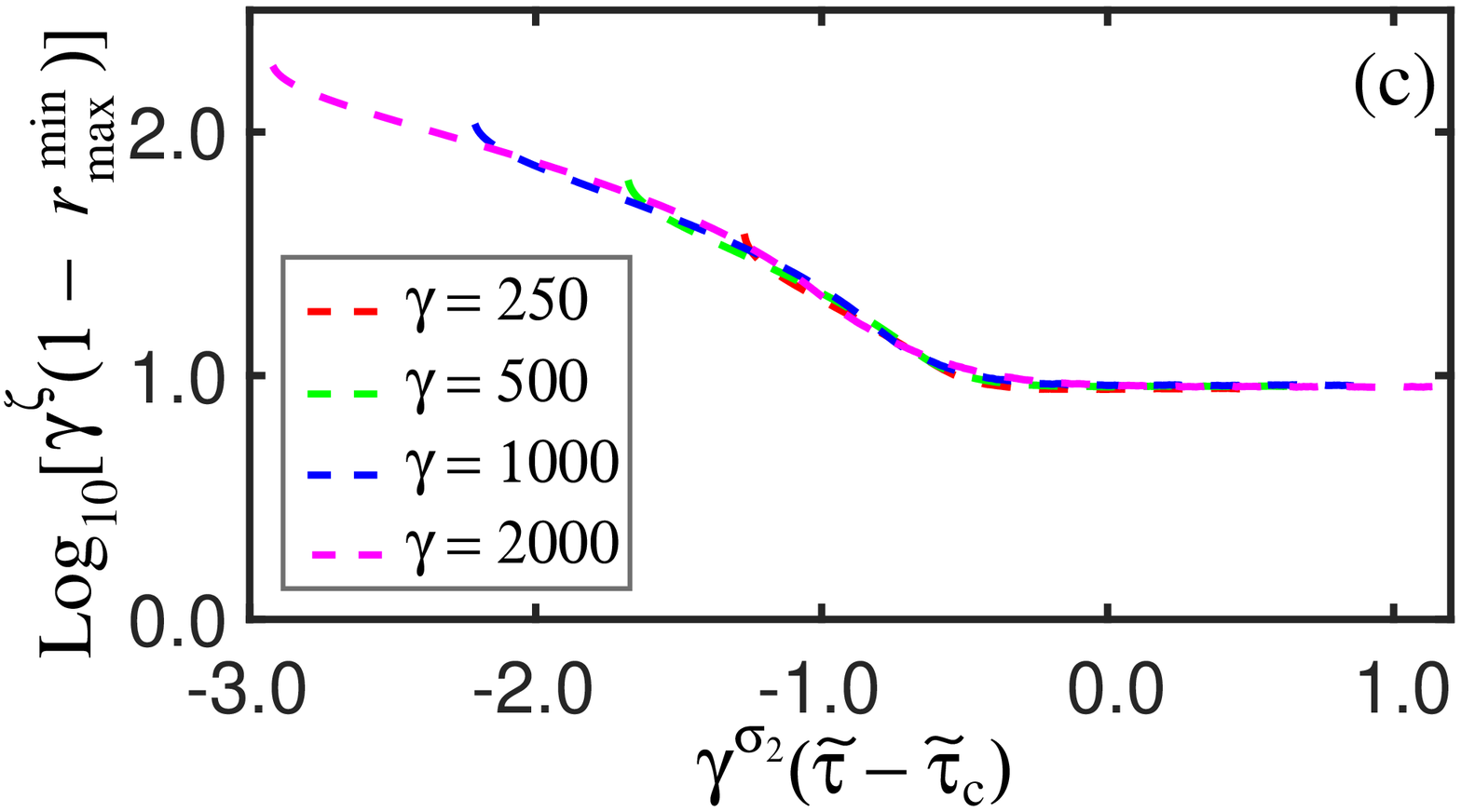}
\caption{\label{fig:Fig11} (Color online) Sphericity complement measures are shown for various $\gamma$ values in panel (a) in the
case of cube shaped protoclasts, with continuous and broken curves representing $1 - r^{\mathrm{min}}_{\mathrm{max}}$ and
$1 - \phi_{\mathrm{3}}$ respectively. Vertical gray and blue lines indicate structural phase transitions involving the removal of 
primordial facets for the former and a reversion of stones to spherical shapes for the latter. Panel (b) and Panel (c) 
display collapses of $1 - \phi_{3}$ and $1 - r_{\mathrm{max}}^{\mathrm{min}}$ onto universal curves.} 
\end{figure}

In Fig.~\ref{fig:Fig11} the complements $1 - \phi_{3}$ and $1 - r_{\mathrm{max}}^{\mathrm{min}}$ are shown for the case of 
cube shaped protoclasts, graphed on a semilogarithmic scale in panel (a); the gray line in the plot corresponds to 
the structural phase transition in which primordial facets disappear.  As in the case of the IPR, both complements are asymptotically flat and 
evenly spaced on the logarithmic ordinate scale, and both measures exhibit singular behavior indicating a structural phase transition which does 
not coincide with the structural transformation in which faces original to the parent cube vanish, but occurs at a later time indicated by the 
vertical blue line in panel (a) of Fig.~\ref{fig:Fig11}.  

In a spirit similar to the case of the IPR, panel (b) and panel (c) of Fig.~\ref{fig:Fig11} show the collapse 
onto universal curves of of $1 - \phi_{3}$ and $1 - r_{\mathrm{max}}^{\mathrm{min}}$ respectively. 
In the case of the former, $\gamma (1 - \phi_{3})$ is plotted with  respect to $\gamma^{\sigma_{1}}(\tilde{\tau} - \tilde{\tau}_{c})$ where 
$\sigma_{1} = 0.23(2)$ and $\tilde{\tau}_{c} = -.137(5)$.  On the other hand, for the latter we achieve a collapse of the 
$r_{\mathrm{max}}^{\mathrm{min}}$ complement measure onto a universal scaling curve by plotting  
$\gamma^{\zeta} (1 - r_{\mathrm{max}}^{\mathrm{min}})$ with respect to $\gamma^{\sigma_{2}} (\tilde{\tau} - \tilde{\tau}_{c})$ where 
the collapse is optimal for $\zeta = 0.81(5)$ and $\sigma_{2} = 0.36(5)$.
In this manner, we see that the stones become smooth even as their primordial facets vanish.

\begin{figure}
\includegraphics[width=.45\textwidth]{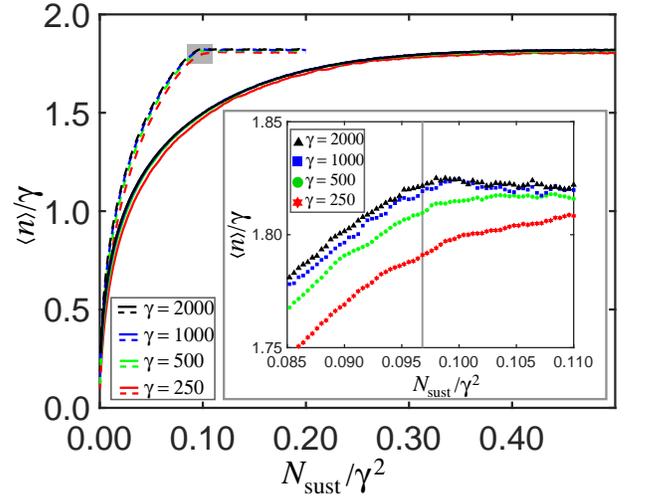}
\caption{\label{fig:Fig12} (Color online) Normalized mean facet numbers, plotted versus $\tilde{\tau} = N_{\mathrm{sust}}/\gamma^{2}$
 are shown for cube shaped protoclasts (broken lines) and
irregular protoclasts (solid lines).  The inset shows a magnified view of the gray rectangular region with symbols
indicating Monte Carlo simulation data points.}
\end{figure}

In Fig.~\ref{fig:Fig12}, normalized mean facet numbers ($\langle n \rangle/\gamma$) are juxtaposed for
cube shaped and regular protoclasts (broken and solid lines respectively), and are well converged with
respect to $\gamma$ in both cases.  In addition, $\langle n \rangle/\gamma$ increases monotonically
with increasing $\tilde{\tau}$ toward the same asymptotic value of 1.82, though the characteristic time constant in
the case of irregular protoclasts exceeds that of the cubic counterparts.  On a qualitative level, while the convergence to
a normalized facet number of 1.82 is gradual for initially irregular fragments, the mean facet number curves for cube shapes protoclasts
saturate for a finite $\tilde{\tau}$ value, quickly becoming level thereafter.  The inset is a magnified view illustrating the
increasing abruptness of change in slope near $\tilde{\tau}_{c}$ with increasing $\gamma$, singular behavior signaling second order phase transition.

\begin{figure}
\includegraphics[width=.45\textwidth]{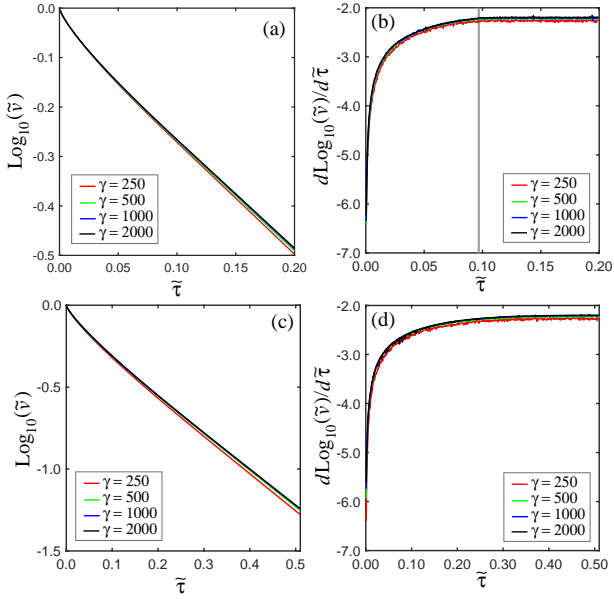}
\caption{\label{fig:Fig13} (Color online) Semilogarithmic plots of volume fraction remaining $\tilde{v}$ with respect to 
$\tilde{\tau}$ as well as the derivatives of the latter are show in panel (a) and panel (b) respectively for cube 
shaped protoclasts; the vertical gray line indicates the structural transition for which primordial facets vanish.  
Similarly, panel (c) and panel (d) show $\tilde{v}$ on a semilogarithmic scale as well as the slope of 
$\log_{10} \tilde{v}$ versus for cohorts of irregular protoclasts.}
\end{figure}

In Fig.~\ref{fig:Fig13}, volume fraction remaining $\tilde{v}$ also reflects singular behavior at the structural transition for 
$\tilde{\tau}_{c} = 0.0968(2)$.  The main graphs in panel (a) and panel (b) of Fig.~\ref{fig:Fig13} show $\tilde{v}$ on a  
semilogarithmic scale, while the inset plots display the sloe of the $\log_{10} (\tilde{v})$ curves with respect to 
$\tilde{\tau} = N_{\mathrm{sust}}/\gamma^{2}$.  In general, as discussed previously, one anticipates the small decrease of 
volume fraction per fracture event to be $d \tilde{v} = -\Delta A^{2}/l$ with $l$ a length scale on the order of 
$\tilde{v}^{1/3}$.  In the regime that stones are worn down to quasi-spherical shapes, one would expect the volume fraction decrement
for a fracture event to be $\Delta  \tilde{v} = -\Delta A^{2}/(12 \pi R)$, exact for the case of a truncated sphere  in the limit that 
$r/R$ (i.e. with r the radius of the new circular facet and $R$ the sphere radius) tends to zero.  Taking  $\Delta A$ to be on the  
order of  $4 \pi R^{2}/\gamma$, we see that the typical volume 
sheared away per normalized time increment $d \tilde{\tau} = \gamma^{-2} \Delta N_{\mathrm{sust}}$ is  
$d \tilde{v} \sim - \tilde{v} d \tilde{\tau}$, which when integrated leads 
to an exponential decay in $\tilde{\tau}$, $\tilde{v} = e^{-\Gamma \tilde{\tau}}$ where $\Gamma$ is a dimensionless constant 
on the order of unity. 

For both cube shaped and irregular protoclasts, the mean remaining volume fraction curves appear to become asymptotically linear, behavior 
highlighted in panel (b) and panel (d) for cubic and irregular parent solids respectively.  Common to both cases, slopes of $\log_{10} \tilde{v}$ 
level out at a common value, which is well converged with respect to the toughness parameter $\gamma$ in both instances.  A salient distinction 
is a discontinuity in the slope of the volume fraction curves  for cube shaped protoclasts not 
replicated in the case of irregular protoclasts.  The latter, in which the slope abruptly becomes constant, signals  pure 
exponential decay of a quasi-spherical shape and also coincides with the facet elimination phase transition indicated  
by the vertical gray line in panel (b) of Fig.~\ref{fig:Fig13}.

\begin{figure}
\includegraphics[width=.45\textwidth]{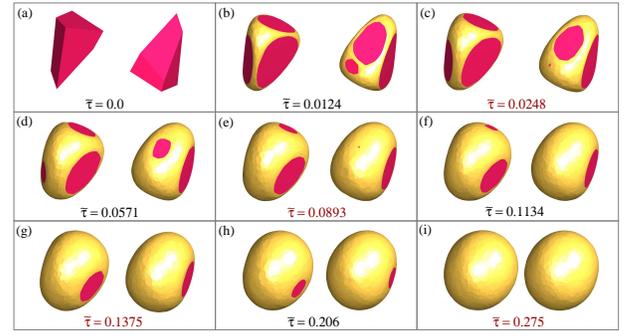}
\caption{\label{fig:Fig14} (Color online) Sequences of structures for an irregular protoclast where 
$\gamma = 2000$.  Images with $\tilde{\tau}$ values in red correspond to transitions involving the 
loss of one or more primordial facets.  Red areas the latter or portions of the latter.} 
\end{figure}

A sharply defined structural phase transition in the case of regular protoclasts such as cubes, even 
for mono-dispersed solids, is the exception rather than the rule, as may be seen in the case of cohorts of identical irregular shapes.   
In this vein, we consider an irregular six sided protoclasts for which the elimination of primordial facets is not simultaneous.  
Instead, the facets of the original protoclast vanish at distinct times $\tau$.  
The parent solid is shown in the upper left corner of Fig.~\ref{fig:Fig14}, illustrating the erosion trajectory 
with intermediate stages (with $\tilde{\tau}$ values in black) interspersed with images corresponding to transitions in 
which one or more primordial facets are removed (with $\tilde{\tau}$ values in red).
Crimson regions are facets original to the parent solid, with the expanding gold areas being material exposed by stochastic fracture events.

\begin{figure}
\includegraphics[width=.45\textwidth]{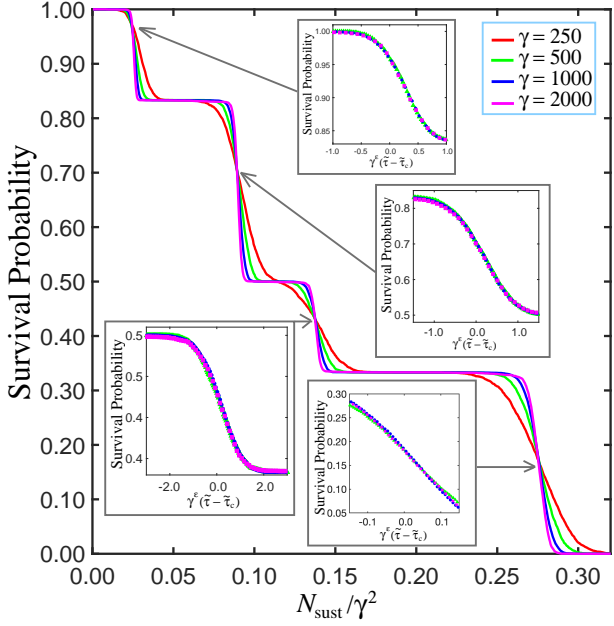}
\caption{\label{fig:Fig15} (Color online) Survival probability for the irregular solid in Fig.~\ref{fig:Fig14}.  Inset 
graphs are data collapses at structural transitions in which one or more facets original to the parent solid are cleaved away.}
\end{figure}

In spite of the absence of a single structural phase transition, the loss of individual primordial facets is accompanied by 
singular behavior in salient variables encountered in the vicinity of a second order phase transition.    
The ensemble averaged facet survival probability shown in Fig.~\ref{fig:Fig15} for various $\gamma$ values 
ranging from $\gamma = 250$ to $\gamma = 2000$ exhibits abrupt transitions signaling the 
loss of individual facets of the original protoclast, with common crossings of the curves for each transition.  While some of 
the latter involve the loss of a single facet, in some cases multiple surfaces from the original protoclast vanish at the same time.  
The first and third transitions [for $\tilde{\tau} = 0.0248(1)$ and $\tilde{\tau} = 0.1375(5)$] are of the former variety, while the second and  
fourth steps downward [for $\tilde{\tau} = 0.0893(1)$ and $\tilde{\tau} = 0.275(1)$] involve the simultaneous disappearance of two 
primordial facets.  The inset graphs are data collapse plots corresponding to the four facet transitions, with the horizontal scale being 
$\gamma^{\varepsilon} (\tilde{\tau} - \tilde{\tau}_{c})$ as in the case of panel (b) of Fig.~\ref{fig:Fig9}.  However, while the data collapses are 
optimized for $\varepsilon = 0.76(5)$ in the case of the first three facet elimination transitions, we find a departure for the transition involving the 
removal of the last two facets of the parent solid where instead $\varepsilon = 0.45(5)$.

\begin{figure}
\includegraphics[width=.45\textwidth]{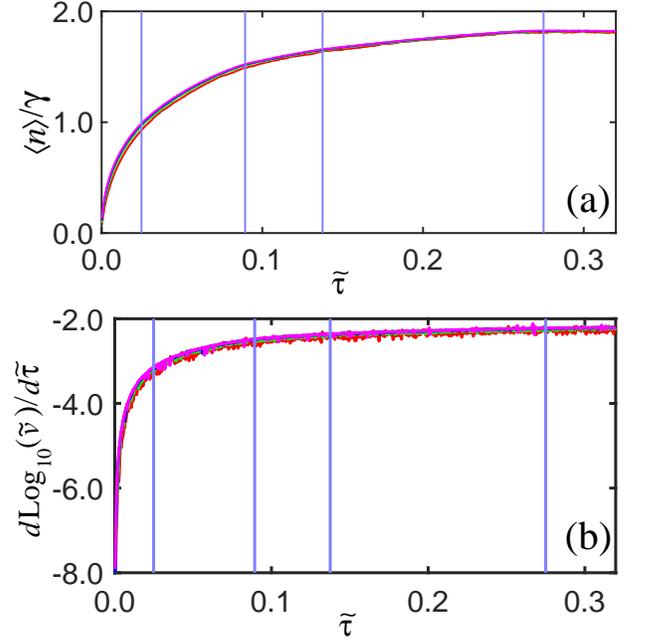}
\caption{\label{fig:Fig16} (Color online) Normalized facet numbers for mono-dispersed irregular protoclasts and 
the slope of the log-log volume curves are depicted in panel (a) and panel (b) respectively.}
\end{figure}

Signatures of the asynchronous facet removal transitions are evident in variables other than $f_{\mathrm{sur}}$, such as the 
mean facet number $\langle n \rangle$ and the slope the logarithm of the mean remaining volume fraction $\tilde{v}$ displayed in panel (a) and panel (b)
of Fig.~\ref{fig:Fig16} for various $\gamma$ values.  In both cases, shift in slopes of the curves are subtler 
than those marking the disappearance of facets original to the parent solid for cubic protoclasts, with slope changes at phase boundaries most pronounced
following the fourth transition in which the final two primordial faces are worn away.

\begin{figure}
\includegraphics[width=.45\textwidth]{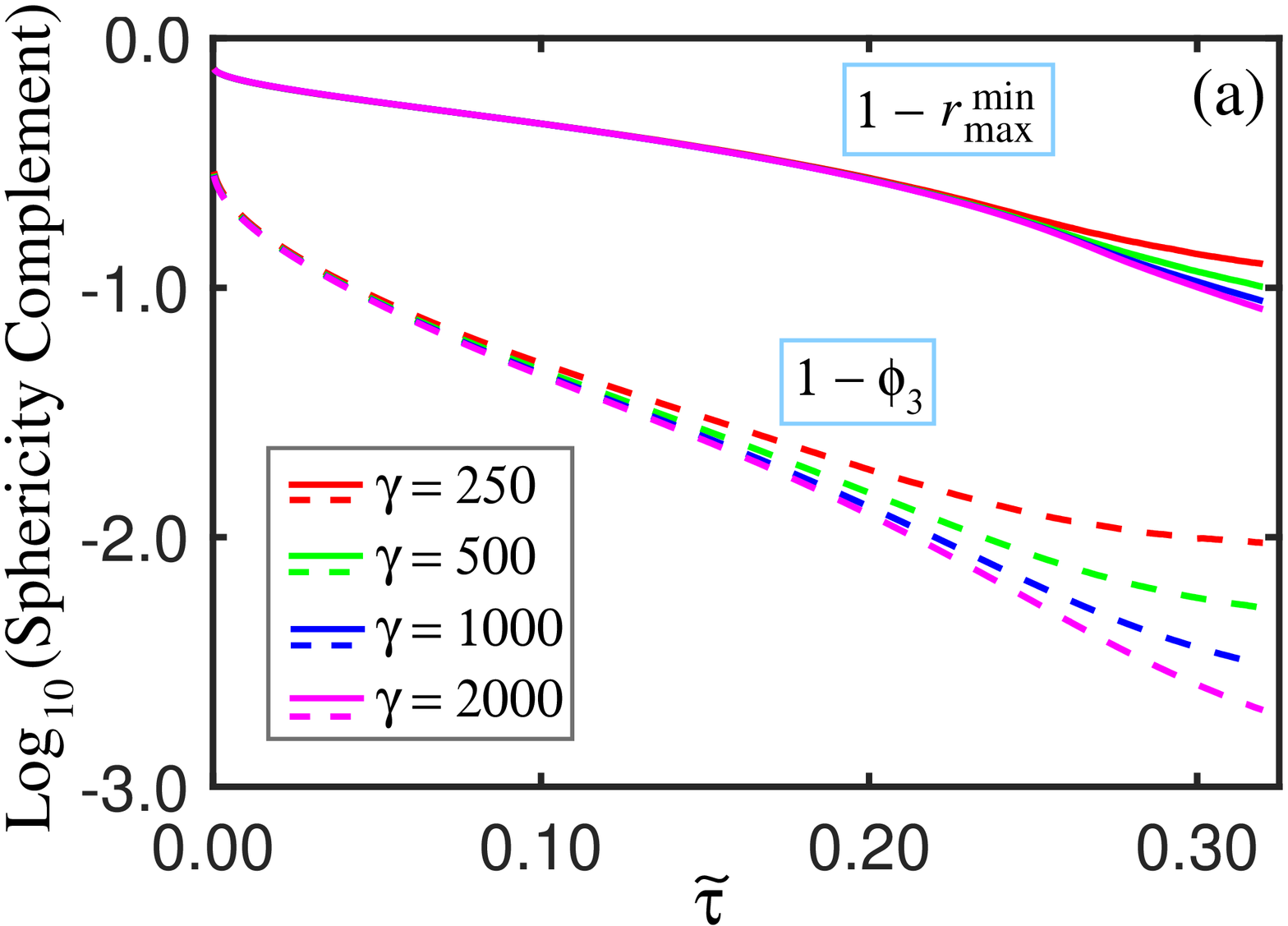}
\includegraphics[width=.45\textwidth]{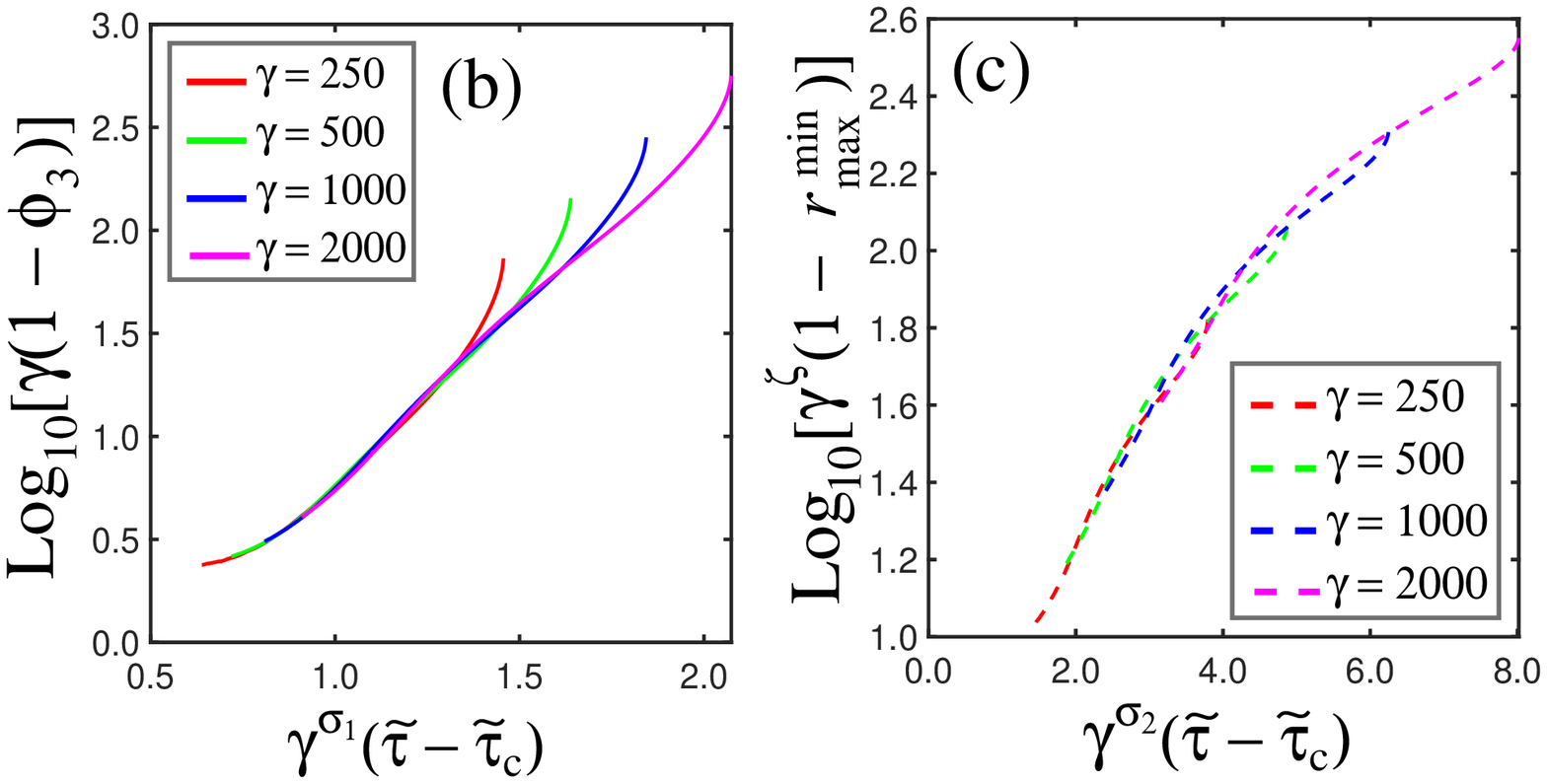}
\caption{\label{fig:Fig17} (Color online) Sphericity complement measures are show for various $\gamma$ values in the
case of mono-dispersed irregular protoclasts, with continuous and broken curves representing $1 - r^{\mathrm{min}}_{\mathrm{max}}$ and
$1 - \phi_{\mathrm{sph}}$ respectively in panel (a).  Panel (b) and panel (c) show collapses of complements $1 - phi_{3}$ and 
$1 - r_{\mathrm{max}}^{\mathrm{min}}$ respectively onto universal scaling curves.}
\end{figure}

Spherical deviation measures $1 - \phi_{3}$ and $1 - r_{\mathrm{max}}^{\mathrm{min}}$ are shown in the main graph of Fig.~\ref{fig:Fig17} on a  
semilogarithmic scale. In both cases, the curves plotted for different $\gamma$ values begin to diverge, in a manner qualitatively similar to the 
case of cube shaped protoclasts, though asymptotically flat regions do not appear on the domain of time scales $\tilde{\tau}$ accessed in the simulation.
Data collapse plots onto universal scaling curves for the sphericity and $r_{\mathrm{max}}^{\mathrm{min}}$ complements are displayed in panels (b) and (c) 
of Fig.~\ref{fig:Fig17}, respectively. 
In the case of the former, $\gamma (1 - \phi_{3})$ is plotted with respect to $\gamma^{\sigma_{1}} (\tilde{\tau} - \tilde{\tau}_{c})$ where
$\sigma_{1} = 0.17(1)$ and $\tilde{\tau}_{c} = 0.57(5)$.  On the other hand, for the latter we plot $\gamma^{\zeta} (1 - r_{\mathrm{max}}^{\mathrm{min}})$ 
with respect to $\gamma^{\sigma_{2}} (\tilde{\tau} - \tilde{\tau}_{c} )$ where $\zeta = 0.81$, $\sigma_{2} = 0.36(5)$, and $\tilde{\tau}_{c} = 0.52(5)$,  
compatible with the transition time obtained for the $\phi_{3}$ complement.
As for cube shaped parent solids, the transition times $\tilde{\tau}_{c}$ occur later than for any of the 
structural transitions in which one or more primordial facets are removed for the irregular mono-dispersed cohort.

\subsubsection{Structural Evolution of Poly-dispersed Cohorts}

\begin{figure}
\includegraphics[width=.45\textwidth]{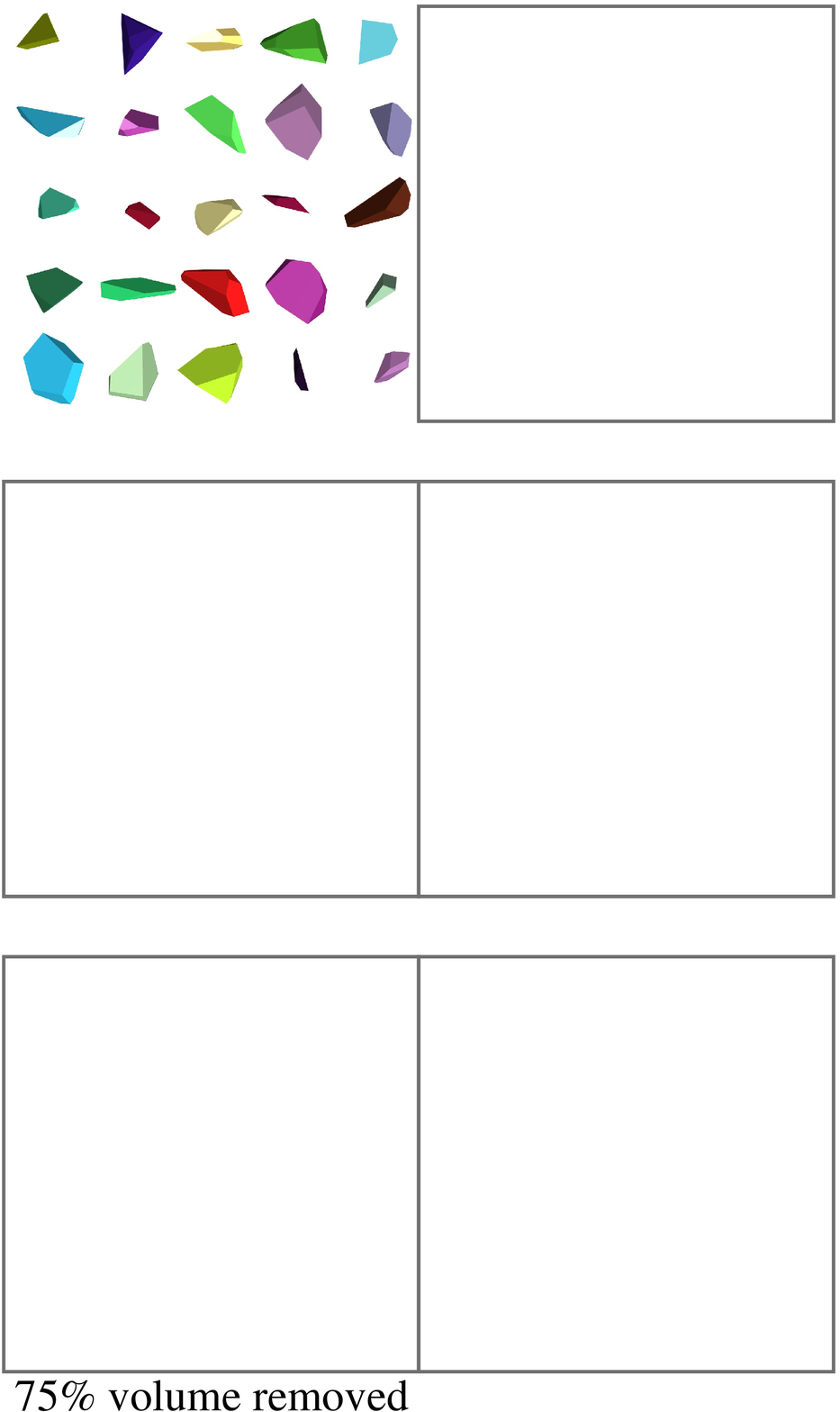}
\caption{\label{fig:Fig18} (Color online) Twenty five irregular protoclasts at various erosion stages}
\end{figure}

Sample stones at various stages of their erosion trajectories are shown in Fig.~\ref{fig:Fig11},
with the irregular protoclasts in the upper left panel.
As in the case of cube shaped protoclasts, individual stones are found to shed
primordial facets, but in an asynchronous manner due to the structural disorder inherent in the
irregular protoclasts, eliminating the possibility of a
well defined structural phase transition.  Thus, variables
exhibit no singular behavior, converging for $\gamma \gg 1$ as in the survival indices plotted in Fig.~\ref{fig:Fig14},
which overlap closely for $\gamma$ ranging from $\gamma = 250$ to $\gamma = 2000$.

\begin{figure}
\includegraphics[width=.45\textwidth]{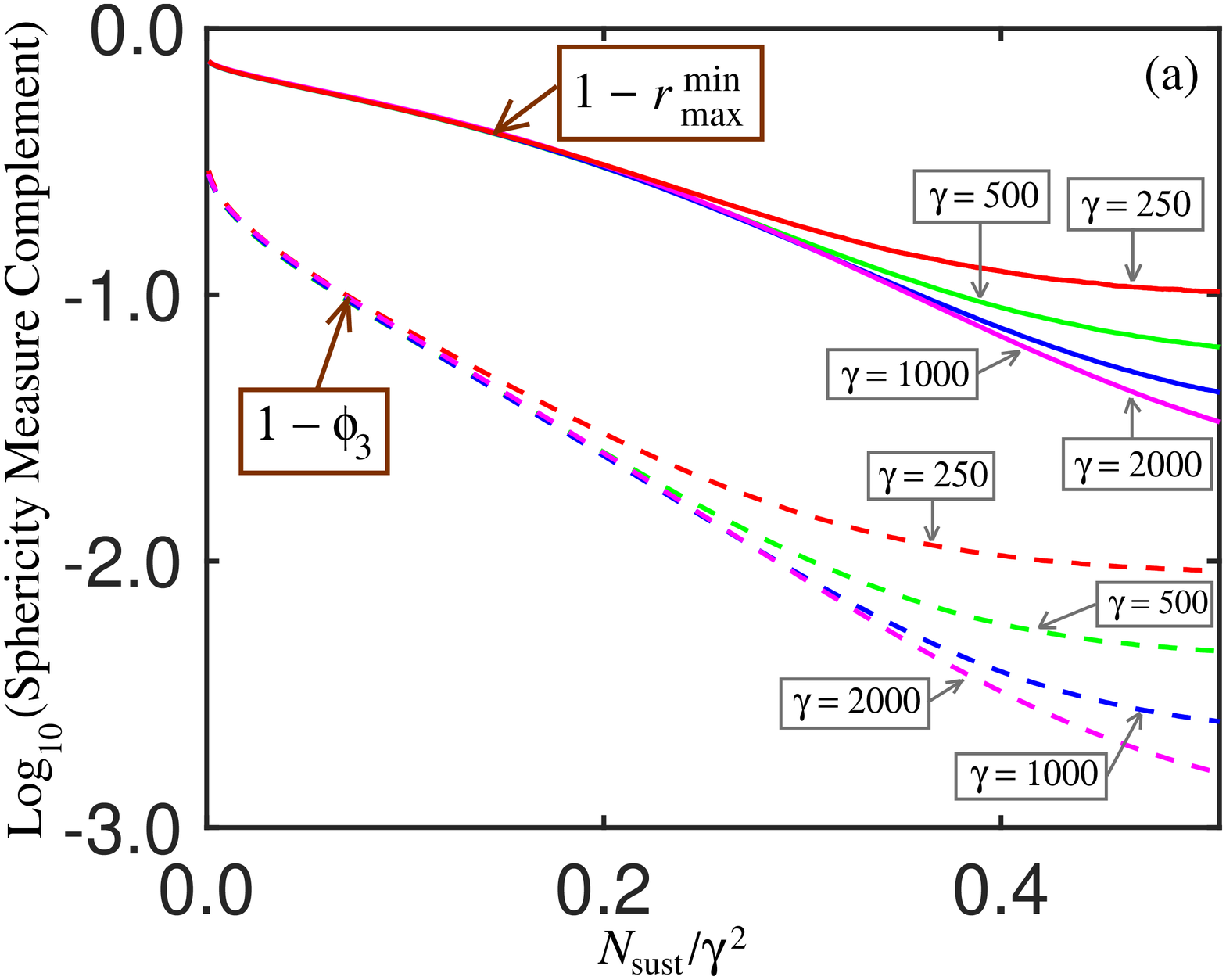}
\includegraphics[width=.45\textwidth]{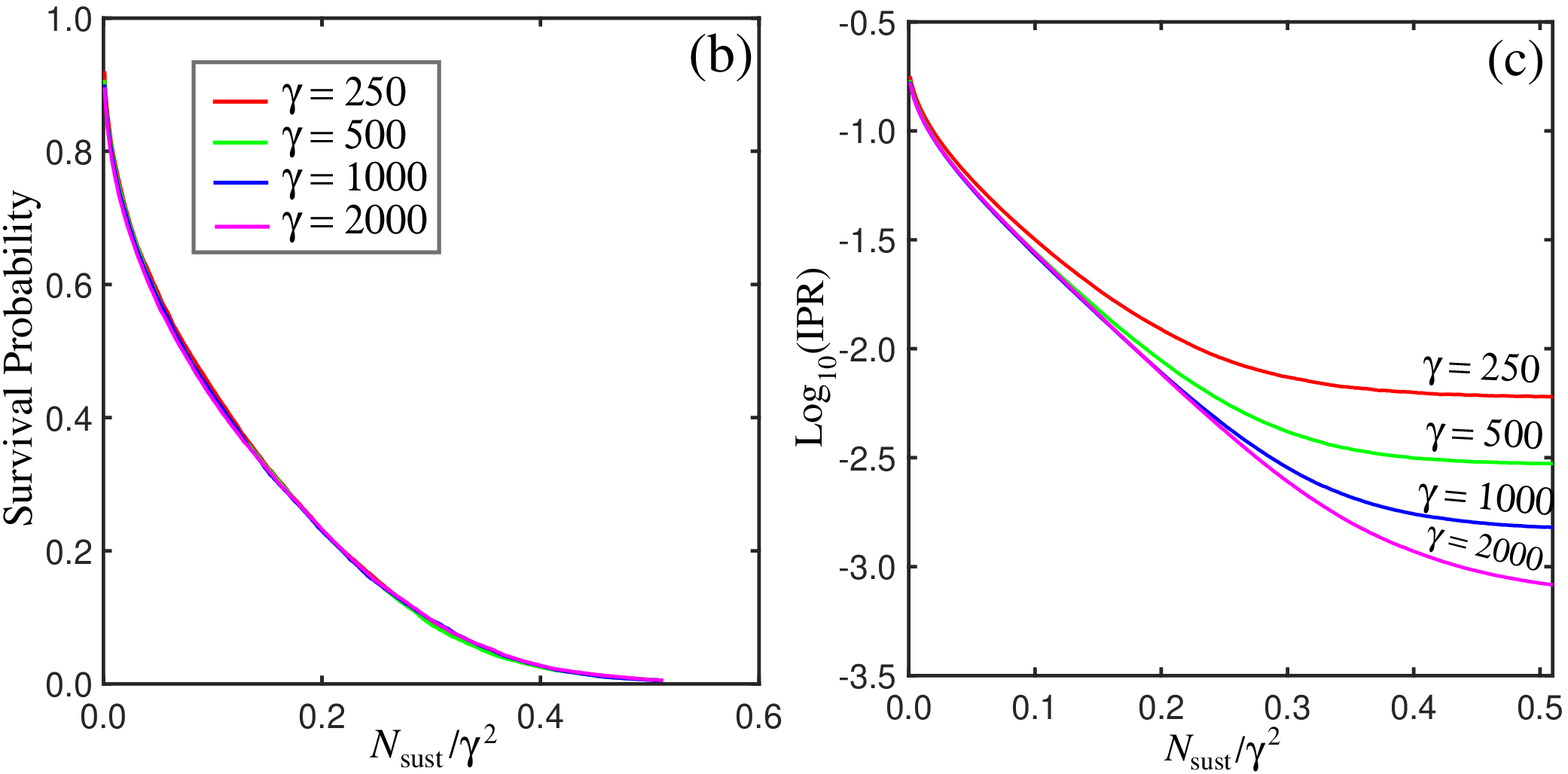}
\caption{\label{fig:Fig19} (Color online) Sphericity complements $1 - \phi_{3}$ and $1 - r^{\mathrm{min}}_{\mathrm{max}}$ 
are shown in panel (a) for irregular protoclast cohorts, while $f_{\mathrm{sur}}$ and the IPR (on a semi-logarithmic scale)
are displayed in panel (b) and panel (c) respectively.}
\end{figure}

Results for the complements $1 - \phi_{3}$ and $1 - f_{\mathrm{max}}^{\mathrm{min}}$, the 
facet survival probability, and the Inverse Participation ratio are shown in panels (a), (b), and (c) 
of Fig.~\ref{fig:Fig19} respectively for irregular protoclasts. In the case of the measures of deviation from a spherical shape in 
panel (a), though the curves diverge, the separation occurs at later and later times with increasing $\gamma$, 
suggesting convergence with respect to the toughness parameter. 
The survival probability shown in panel (b) of Fig.~\ref{fig:Fig19} evidently converges rapidly with increasing $\gamma$ when plotted
with respect to $N_{\mathrm{sust}}/\gamma^{2}$ with little difference among the curves for all $\gamma$ values shown.
In contrast to the case of cube shaped parent stones, IPR curves for irregularly shaped protoclasts 
converge with increasing $\gamma$, and there is no finite value of $\tilde{\tau}$ where the participation ratio
decays to zero in the large $\gamma$ limit.  This characteristic is compatible with primordial facet 
removal events being spread over the entire $\tilde{\tau}$ domain due to the strong structural disorder in the 
cohort of poly-dispersed irregular protoclasts.

\begin{figure}
\includegraphics[width=.45\textwidth]{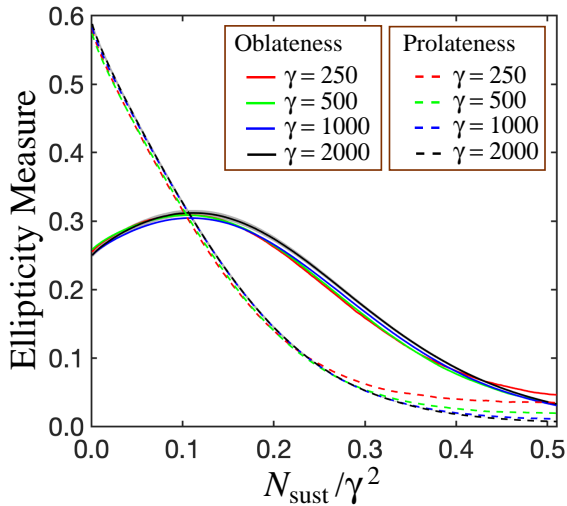}
\caption{\label{fig:Fig20} (Color online) Ellipticity measures plotted versus $\tilde{\tau}$ 
for various $\gamma$ values, with solid and broken lines representing oblateness and prolateness respectively. Gray regions 
about $\gamma = 2000$ curves indicate Monte Carlo statistical error.}
\end{figure}

Oblateness and prolateness measures $\psi_{\mathrm{P}}$ and 
$\psi_{\mathrm{O}}$, as described in Section II, are displayed in Fig.~\ref{fig:Fig20} for a range of $\gamma$ values. The 
prolateness measure initially exceeds the oblateness measure, but eventually decreases sharply and falls below $\psi_{\mathrm{O}}$ with a 
continued monotonic decrease thereafter.  
A study of strongly disordered fragments~\cite{Domokos4} generated stochastically found a prevalence of prolate fragments in cohorts 
generated with a variety of techniques.
On the other hand, the disorder averaged oblateness measure actually increases initially, peaking 
when approximately half of the stone's original volume has been chipped away.  The variance in the $\psi_{\mathrm{O}}$ curves, while greater than 
for the $\psi_{\mathrm{P}}$ traces, is comparable to the Monte Carlo statistical error shade in gray in the case of the $\gamma = 2000$ curve,
and thus both the oblateness and prolateness measures may be considered to be converged with respect to the toughness parameter $\gamma$.
That $\psi_{\mathrm{O}}$ peaks even as $\psi_{\mathrm{P}}$ continues to decrease could predispose stones to long term oblate shapes, though much 
of the enhancement and maintenance of an oblate profile, not addressed in this study, 
may be due to specific nature of fluvial motion along a stream or river bed~\cite{Oblate1}.
 
\begin{figure}
\includegraphics[width=.45\textwidth]{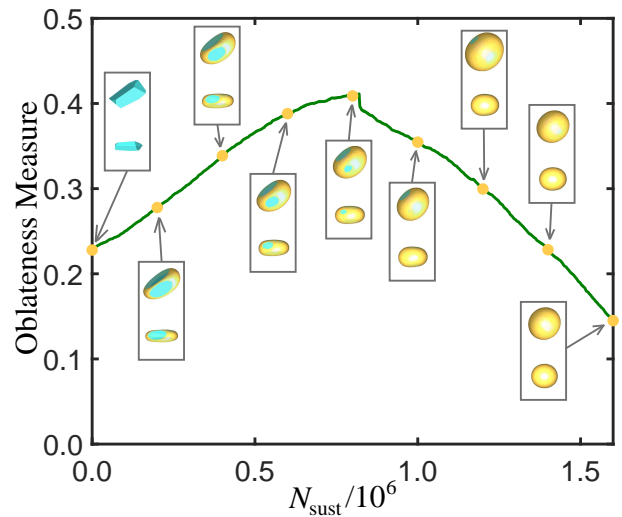}
\caption{\label{fig:Fig21} (Color online) Oblateness measure $\psi_{\mathrm{O}}$ for an individual case showing 
non-monotonic behavior in the oblateness.  The solid curve is the oblateness measure plotted with respect to time, and 
rectangular gray boxes enclose images of the stones at specific stages.  The lower image in each case is an edge-on
view of the same stone.}
\end{figure}

To gain an understanding on an intuitive level as to why the time dependent oblateness measure is non-monotonic in some cases, 
we exhibit $\psi_{\mathrm{O}}$ as well as images of the stone at various stages in the erosion trajectory.  The protoclast for the example 
shown in Fig.~\ref{fig:Fig21} begins as 
a blade-like shape, narrow but also thin. As fractures begin to carve away volume, the stone becomes thicker (as may be seen in the 
edge-on views), but appears to lose material from the extremes along the longer axis at an even greater rate.  In combination, these trends 
contribute to an initial rise in oblateness (with the shape remaining relatively thin while becoming less oblong), ultimately peaking an declining  
as the stone ultimately is rounded into a spherical shape.  

\section{Non Steady State Erosion Scenarios}

\begin{figure}
\includegraphics[width=.45\textwidth]{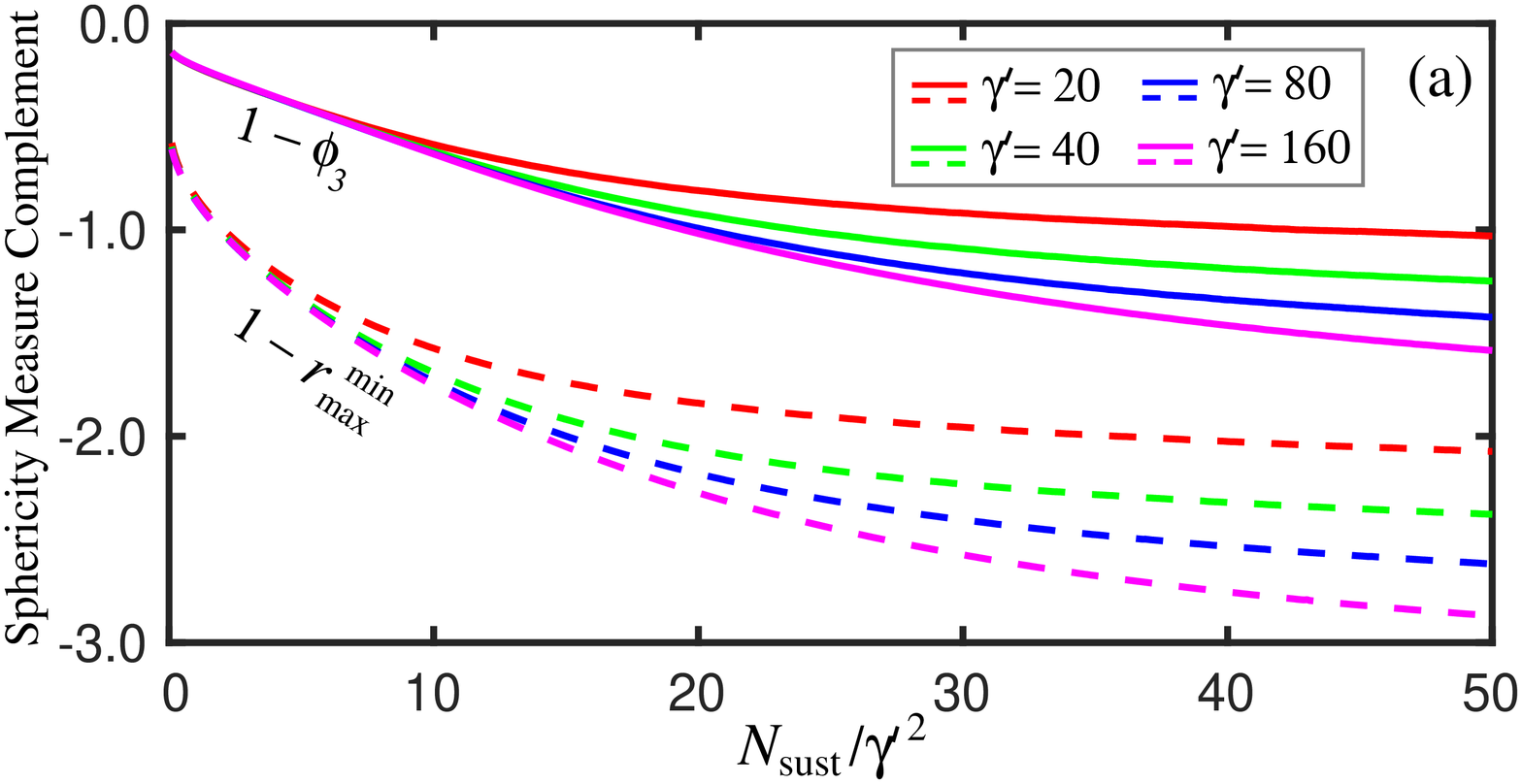}
\includegraphics[width=.45\textwidth]{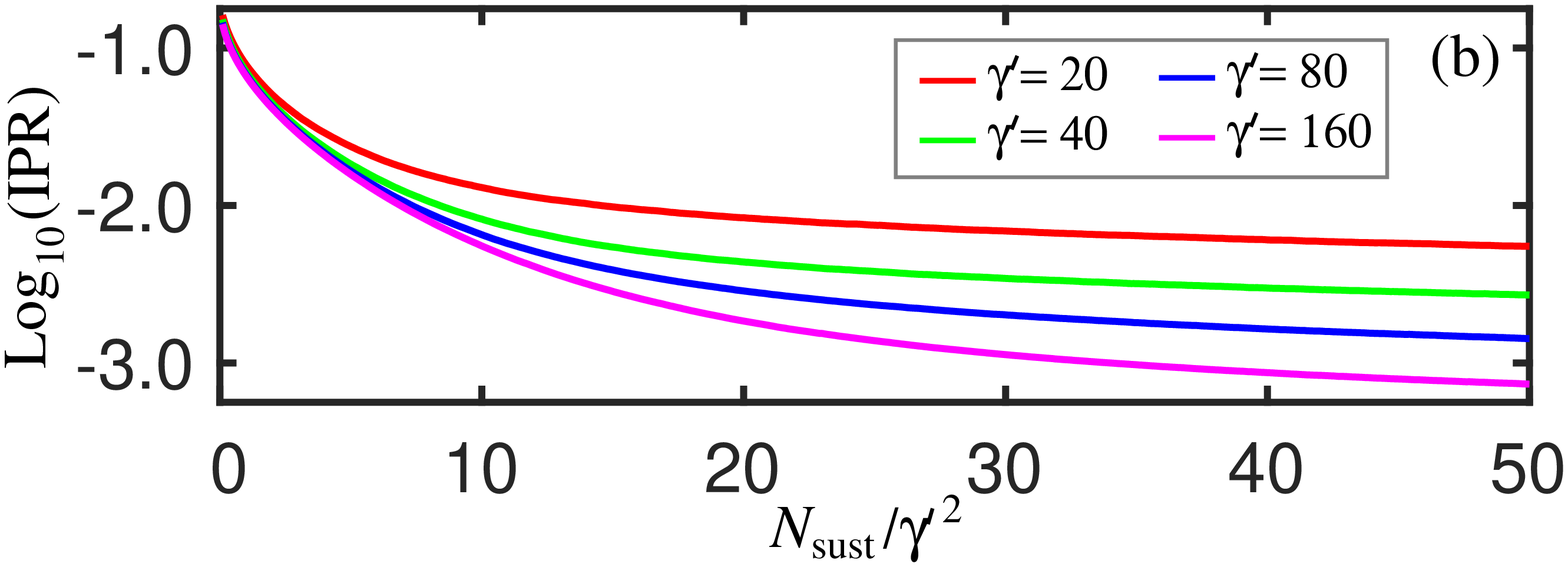}
\includegraphics[width=.45\textwidth]{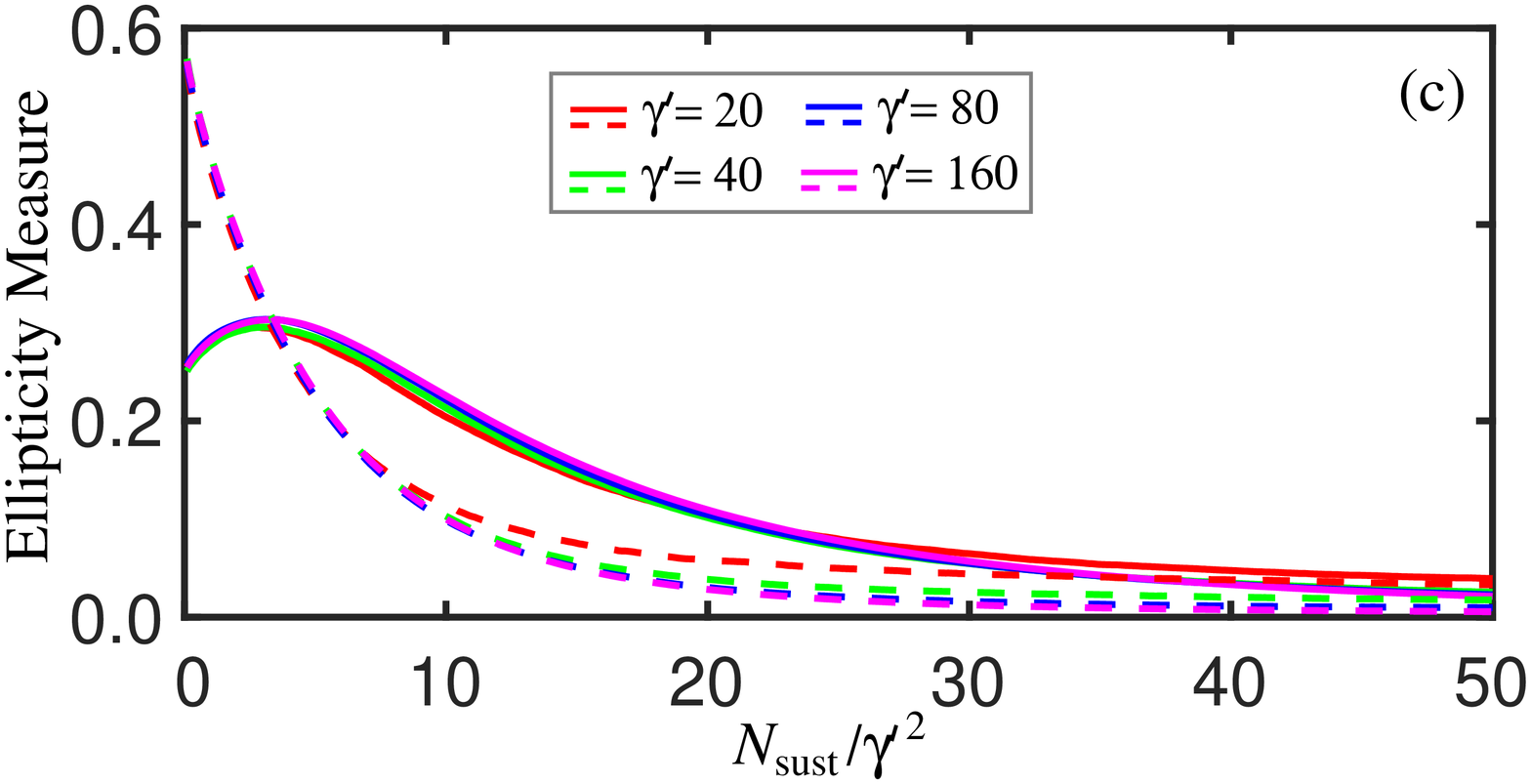}
\includegraphics[width=.45\textwidth]{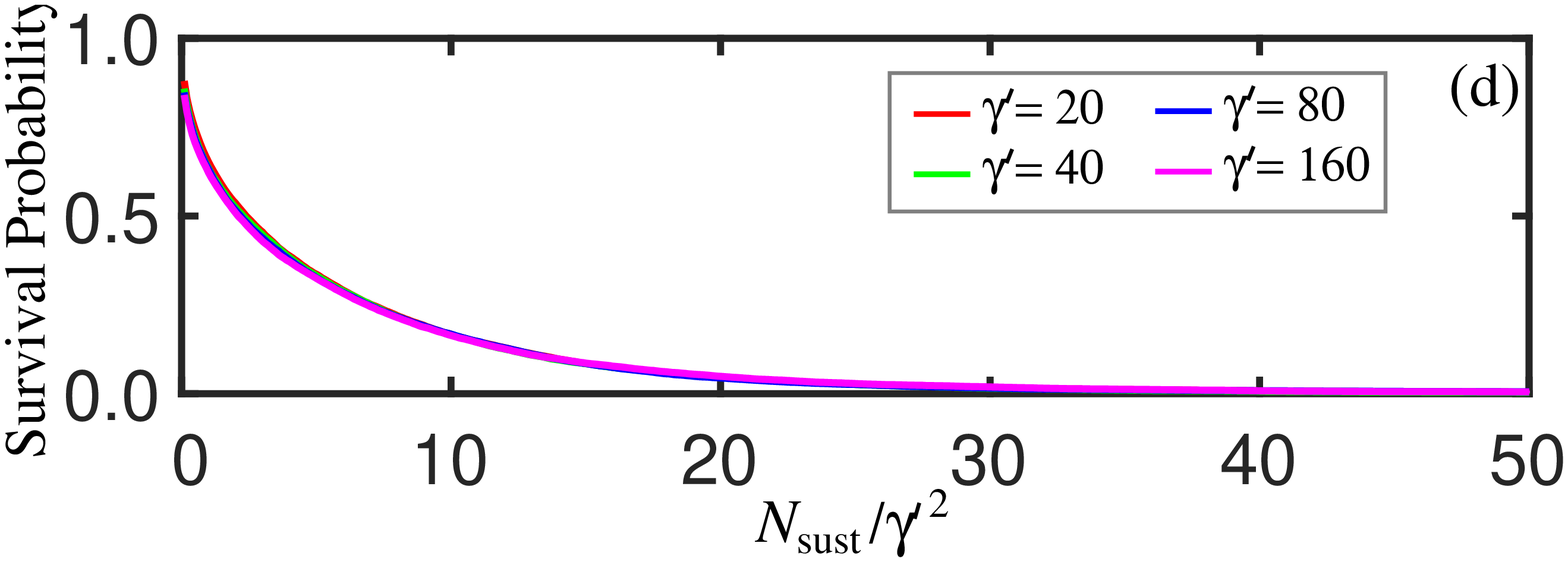}
\caption{\label{fig:Fig22} (Color online) Observables plotted with respect to $N_{\mathrm{sust}}/(\gamma^{'})^{2}$ 
for $1 - r_{\mathrm{max}}^{\mathrm{min}}$ and $1 - \phi_{3}$ in panel (a), the IPR in panel (b), the oblateness and prolateness 
in panel (c), and the primordial facet survival probability in panel (d).} 
\end{figure}

Although we discuss a variety of non-steady state schemes, choose the constant 
velocity scenario (e.g. for stones borne along at constant speed in a stream or river)
for direct Monte Carlo simulations.  With kinetic energy scaling as $\frac{1}{2} m v^{2}$, and 
$m = \rho V$ in the case of stones of uniform composition, typical energy would be proportional to the 
volume.  That the $\alpha = 1$ exponent for the fixed velocity scenario exceeds the critical $\alpha_{c} = 2/3$ 
for the relative area scheme
implies an ever diminishing relative area with time for typical sustained slices and a concomitant 
rise in the mean number of facets.  Nevertheless, although the relative area and fixed 
velocity scenarios are distinct, the universal dependence of stone profiles on fractional remaining volume~\cite{Domokos1}  
dictates that key structural milestones occur at finite values of the reduced time $\tilde{\tau}$, 
borne out in the case of the fixed velocity scheme.

Fixed velocity erosion trajectories are calculated for irregular protoclasts with 2000 distinct realizations of disorder. 
A range of toughness parameter values from $\gamma^{'} = 20$ to $\gamma^{'} = 160$ 
are considered with the mean number of facets $\langle n \rangle$ exceeding 4000 for the latter.
Representative examples of relevant variables are shown in Fig.~\ref{fig:Fig22} with the sphericity complements $1 - \phi_{3}$ and 
$1 - r_{\mathrm{max}}^{\mathrm{min}}$ and the IPR shown in panel (a) and (b) while 
oblateness and prolateness measures are shown in panel (c) with the primordial facet survival probability $f_{\mathrm{sur}}$  plotted in 
panel (d).  As in the case of the time dependent relative area results, $f_{\mathrm{sur}}$ and ellipticity measures are converged with 
respect to the $\gamma^{'}$ parameter with minor discrepancies for the oblateness measure near its maximum likely due to Monte 
Carlo statistical error. Quantities such as measures of deviation from a spherical shape and the IPR do diverge on the semilogarithmic 
scale, but at later times with increasing $\gamma$ suggesting convergence with respect to the latter.
In a qualitative sense, in comparison to results in the case of the relative area scenario, 
fixed velocity variables evolve more rapidly initially with structural change slowing as sustained slices accumulate. This trend 
is compatible with the typical area of freshly exposed facets decreasing relative to the total area with decreasing remaining volume 
fraction $\tilde{v}$.  

\begin{figure}
\includegraphics[width=.45\textwidth]{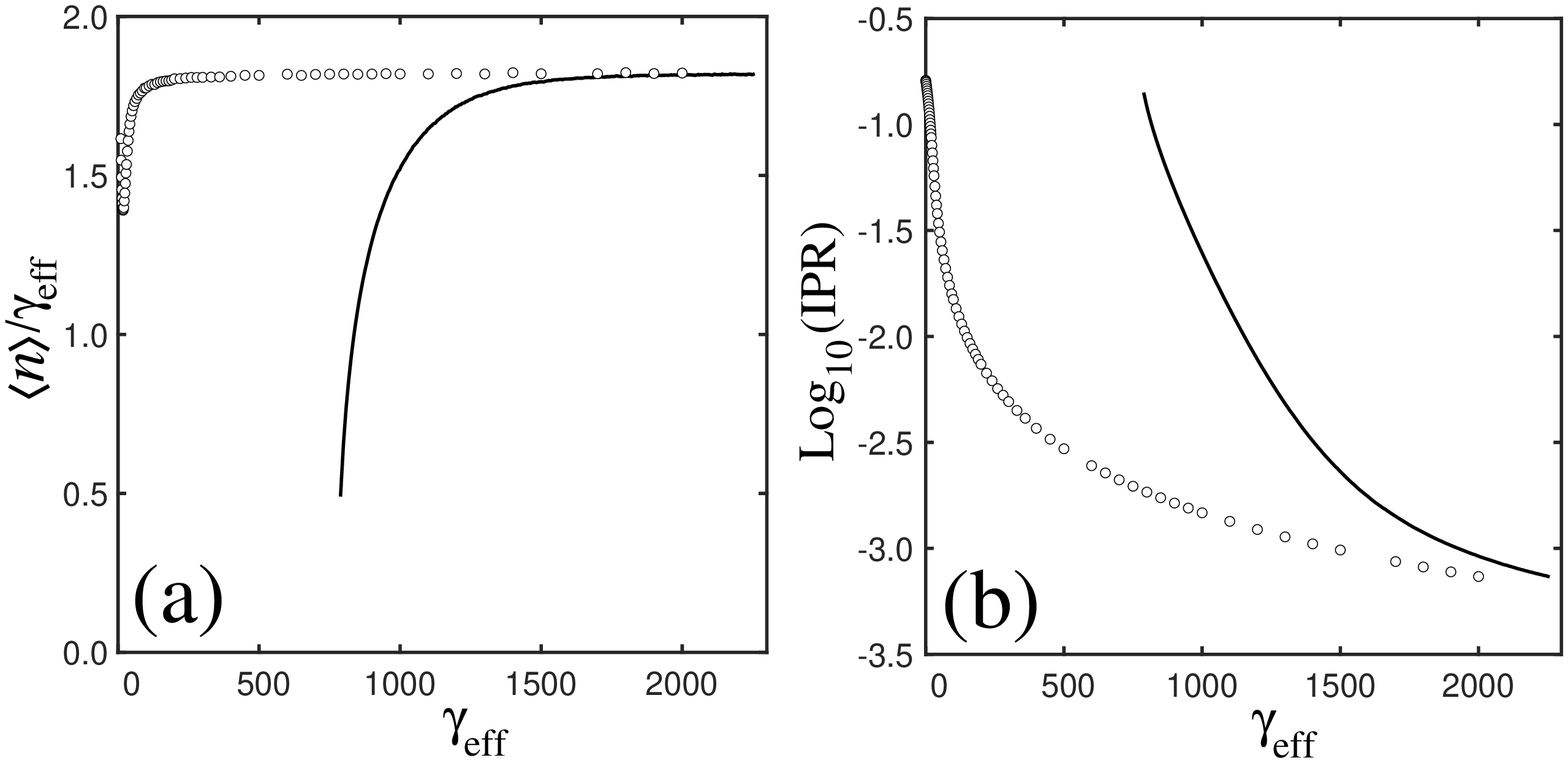}
\includegraphics[width=.45\textwidth]{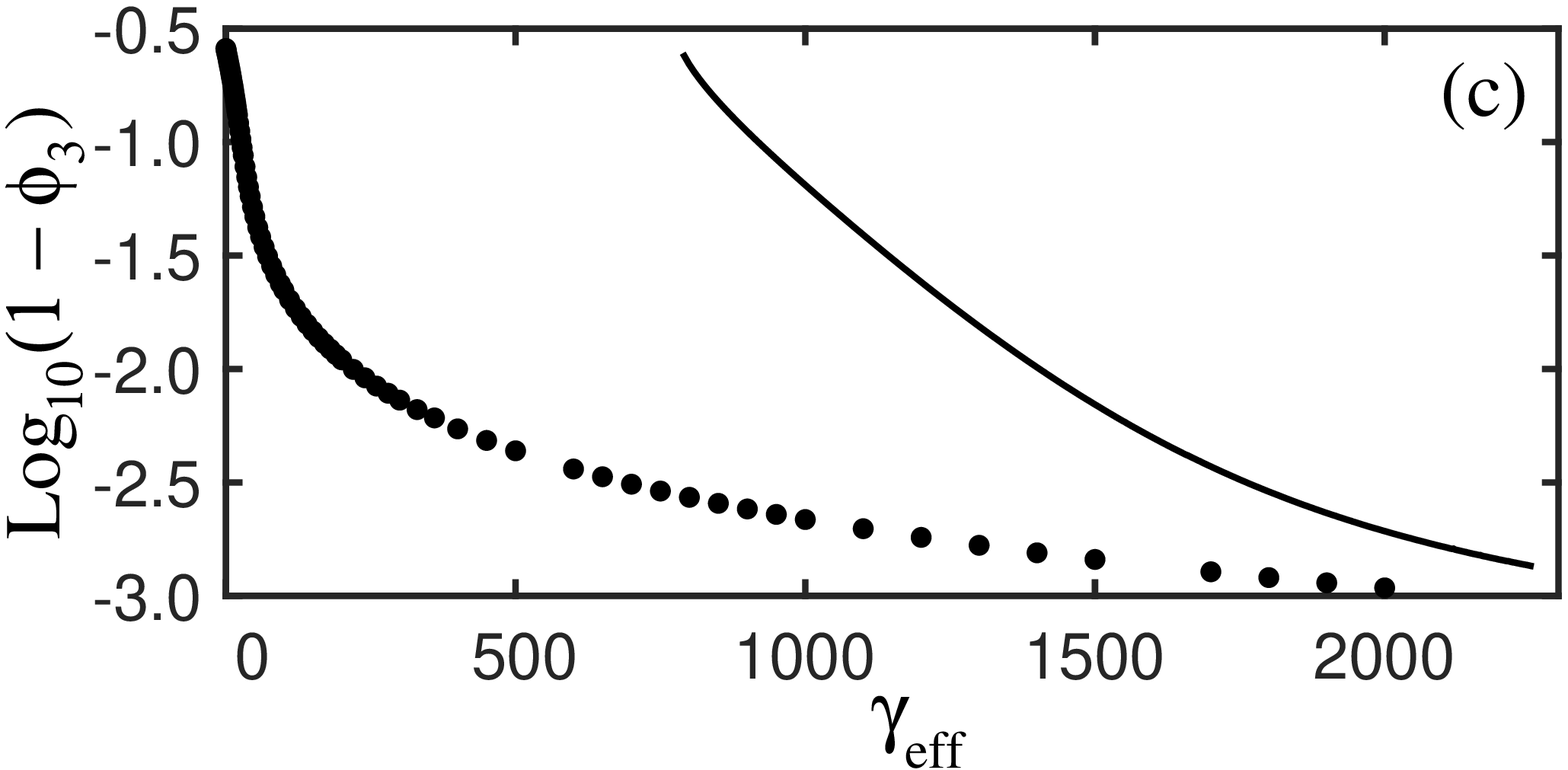}
\includegraphics[width=.45\textwidth]{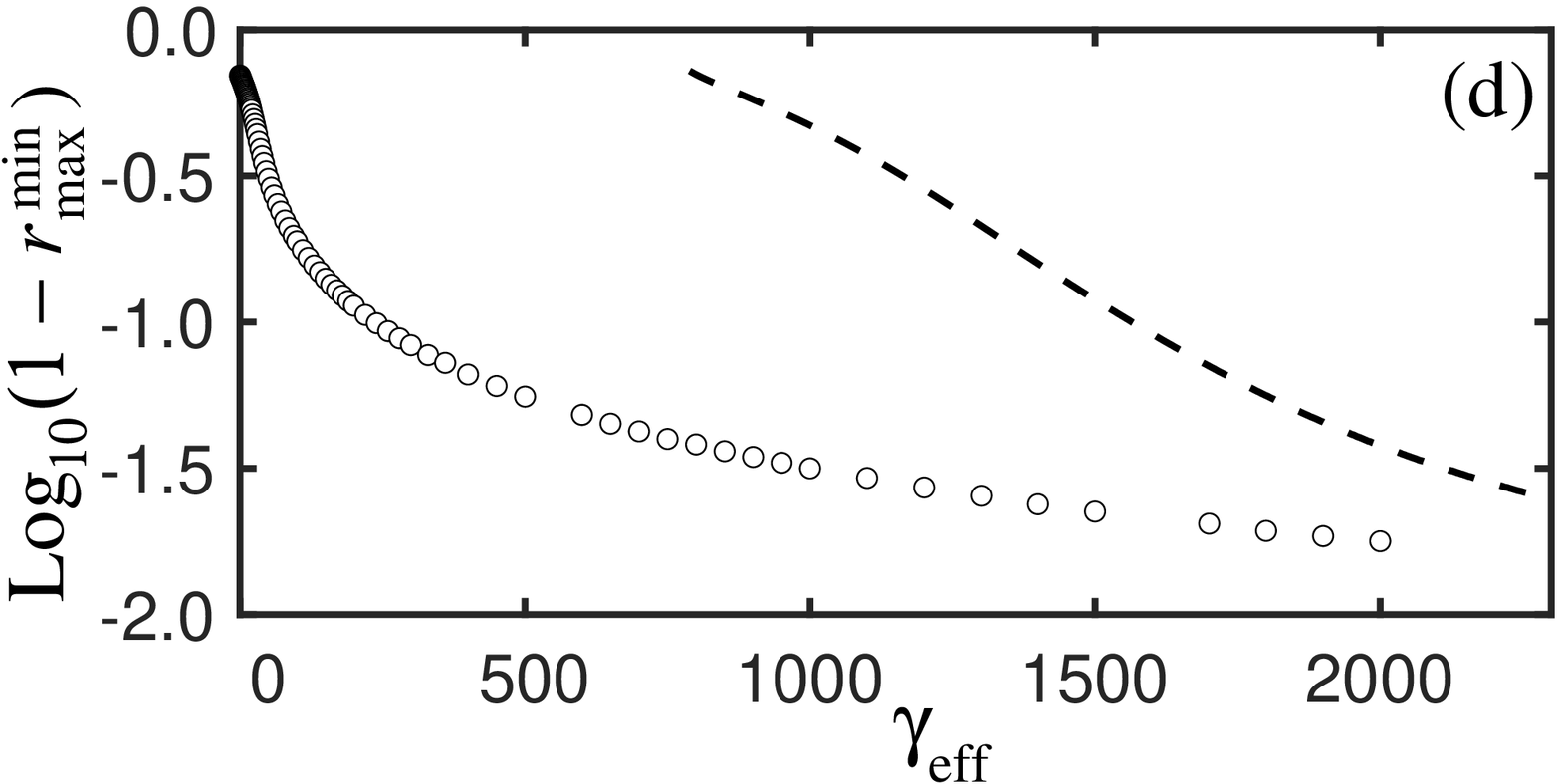}
\caption{\label{fig:Fig23} (color online) Effective Gamma plots in the fixed velocity scenario are displayed for
the mean facet number $\langle n \rangle$ and IPR in panel (a) and panel (b), and spherical complements $1 - \phi_{3}$ and
$1 - r_{\mathrm{max}}^{\mathrm{min}}$ in panel (c) and panel (d) respectively.  Symbols represent equilibrated steady state
systems, while broken and solid lines are simulation data in the fixed velocity scenario.}
\end{figure}

An additional merit of discussing the steady state scenario in the equilibrated regime is its relevance to 
non-steady state scenarios after a sufficient number of fractures have accumulated.  We define an instantaneous or 
``effective'' $\gamma$ with $\gamma_{\mathrm{eff}} = A_{\Sigma}/A_{\textrm{char}}$ where $A_{\textrm{char}}$ is the characteristic 
area exposed by a fracture event obtained by setting the exponent in the prospective slice acceptance probability to 1 and solving 
for the area.  
In the case of the fixed velocity erosion scheme, one has 
$\gamma_{\mathrm{eff}} = \tilde{v}^{-\frac{1}{3}} (6 \pi^{\frac{1}{2}})^{-1} \gamma^{'}$, which 
increases with decreasing remaining volume fraction.   
For representative variables of interest in Fig.~\ref{fig:Fig23}, we show on the same plots equilibrated observables from 
steady state calculations and fixed velocity simulation results (solid or broken curves) plotted with respect to $\gamma_{\mathrm{eff}}$.  
The mean number of sites normalized with respect to  $\gamma_{\mathrm{eff}}$, shown in panel (a), ultimately saturates at the 
large $\gamma$ equilibrium value of $1.82$.  The IPR and complements $1 - \phi_{3}$ 
and $1 - r_{\mathrm{max}}^{\mathrm{min}}$ obtained from fixed velocity scheme 
simulations decrease monotonically with time in panels (b), panel (c), and panel (d) of Fig.~\ref{fig:Fig23},
converging with steady state equilibrium results (open or filled circles). The merging of the fixed velocity simulation and equilibrium results 
suggests that even in non-steady state situations, stones reach a condition of quasi-equilibrium, in the sense that variables correspond closely 
with the steady-state counterparts calculated for $\gamma_{\mathrm{eff}}$ after enough time has elapsed.

\subsection{General Scenario Variables from Relative Area Results}

\begin{figure}
\includegraphics[width=.45\textwidth]{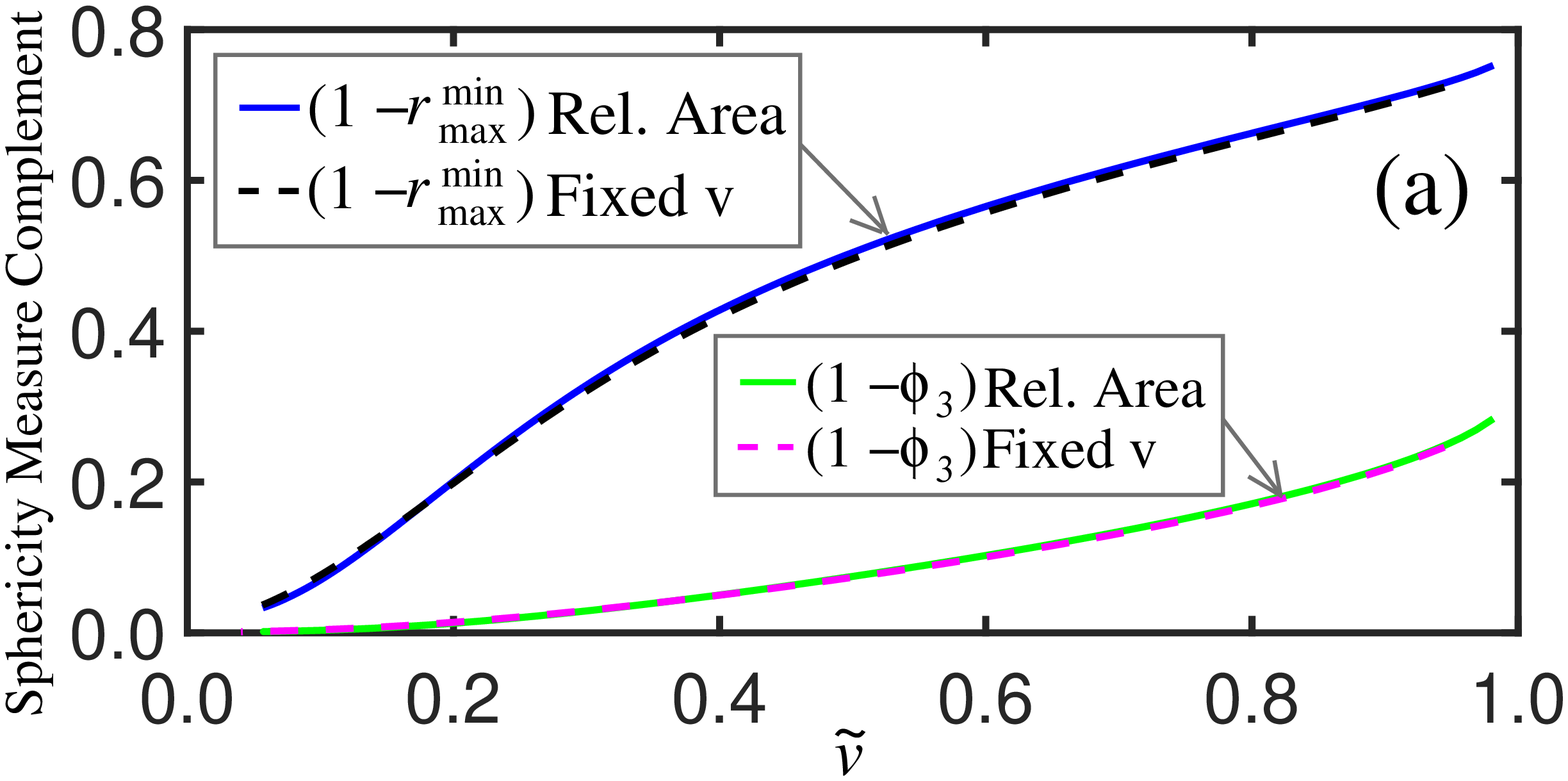}
\includegraphics[width=.45\textwidth]{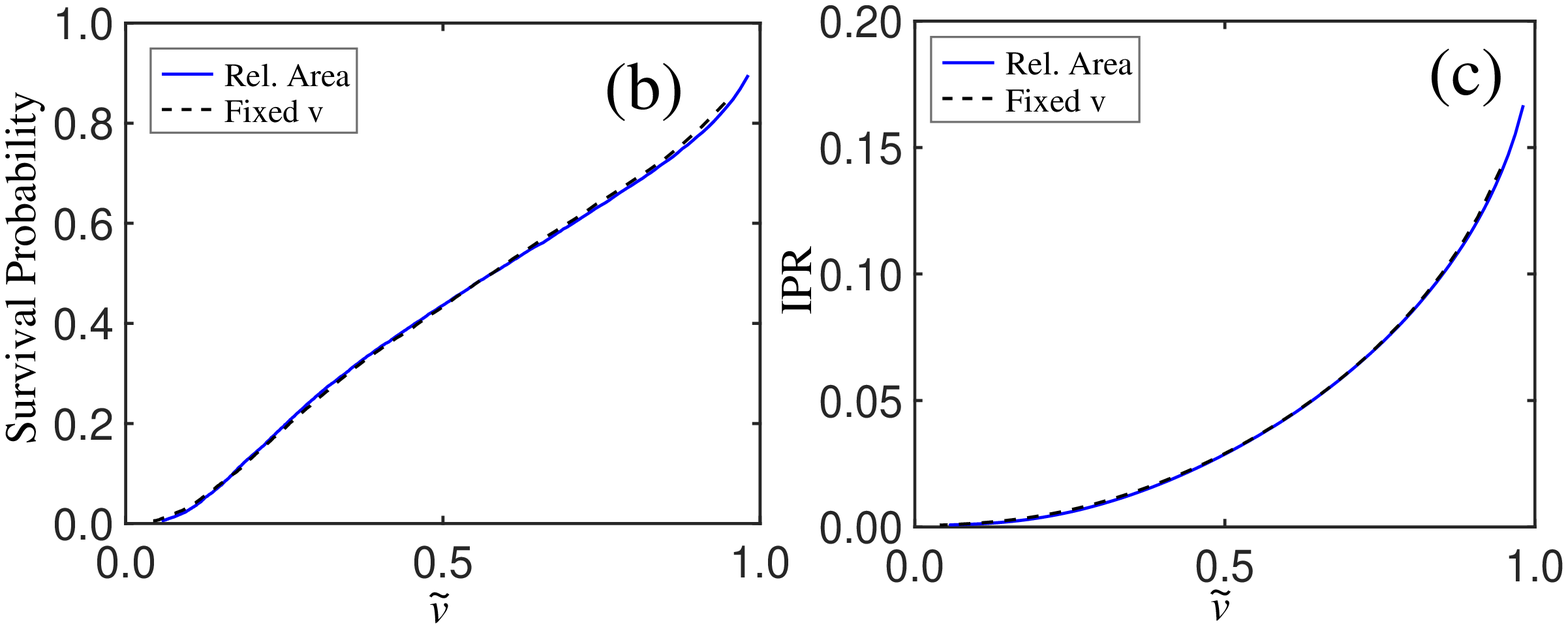}
\includegraphics[width=.45\textwidth]{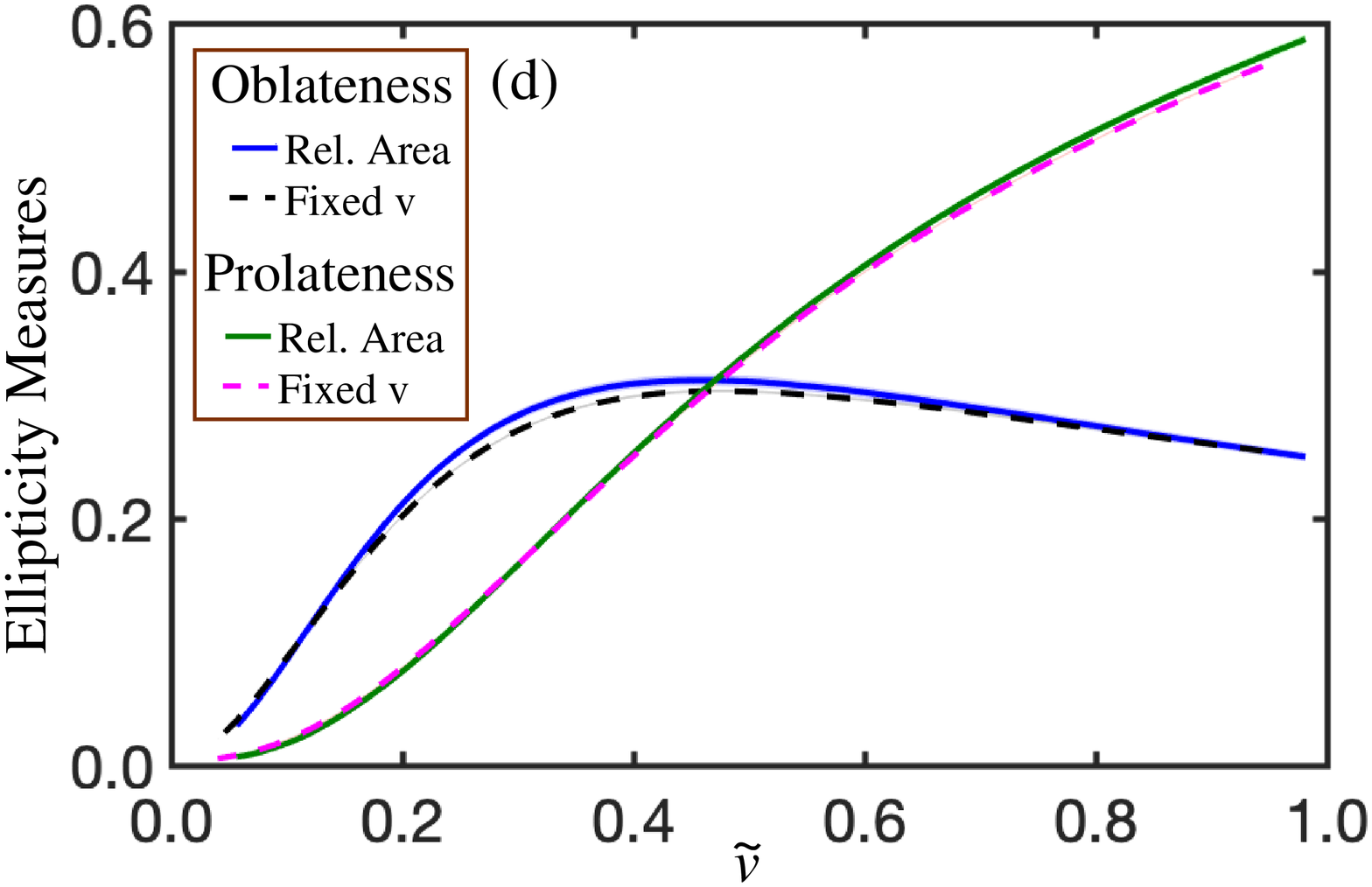}
\caption{\label{fig:Fig24} (Color online) Variables for relative area and fixed velocity 
scenarios plotted with respect to volume fraction $\tilde{v}$.}
\end{figure}

Equivalent milestones occurring at the same volume fractions remaining 
independent of the model (e.g. the oblateness maxima near the halfway point in the erosion of 
volume) is an important aspect of the deterministic time evolution of structures in spite
of the stochastic nature of the collisional erosion process; in more succinct terms, 
graphs of relevant observables should coincide when plotted versus $\tilde{v}$ even for distinct models, 
a phenomenon which we see in Figure~\ref{fig:Fig24} for sample observables of interest.  
In panel (a), global measures of departure from a perfect spherical shape are plotted with respect to  
$\tilde{v}$, with the solid curve gleaned from relative area calculations and the broken trace corresponding to 
the fixed velocity scenario.  Panel (b) and (c), depicting the mean facet survival probability and the 
Inverse Participation Ratio also show very close agreement among the relative area and constant velocity 
results when plotted versus the volume fraction remaining.  Finally, measures of ellipticity are displayed in 
panel (d) of Fig.~\ref{fig:Fig24}, with close agreement for both the prolateness and oblateness measures.
This quantitative agreement of result for distinct scenarios when plotted with respect to the remaining volume fraction
has been mentioned previously~\cite{Domokos1}.

The relative area scenario (i.e. $\alpha = \alpha_{c} = 2/3$)  considered in the preceding section provides an
avenue for predicting the time evolution of salient observables in a broad set of cases 
in which the number of facets does not ultimately saturate at a finite value, but continues to 
rise monotonically.   
The universal dependence of variables on the mean volume fraction $\tilde{v}$  for steady state 
and non-steady state scenarios (e.g. the constant velocity scheme) would permit a prediction of the 
time dependence of variables in a wide variety of non-steady state scenarios if one knew 
the relationship of $N_{\mathrm{sust}}$ and $\tilde{v}$ for the scheme under consideration.  
We argue here and validate with large scale simulations involving irregular 
protoclasts that this mapping of $\tilde{v}$ onto $N_{\mathrm{sust}}$  in greater 
generality may be determined from steady state results as long 
as $\gamma \gg 1$.  

With the characteristic mean facet area $A_{\mathrm{char}}$  being on the order of $A_{\mathrm{char}} = \kappa \gamma^{-1} \tilde{v}^{2/3}$, where $\kappa \equiv
(36 \pi^{2})^{1/3}$, the volume cleaved away
by a sustained slice scales as $A_{\mathrm{char}}^{2}/\tilde{r}$, where $\tilde{r}$ is as discussed earlier a length scale
related to the local radii of curvature, which we take to be on the order of $\tilde{v}^{1/3}$, the cube root of the volume fraction.
We obtain an equality with a volume dependent function $f(\tilde{v})$,  yielding for the mean volume decrement
$\langle \Delta \tilde{v} \rangle = - \kappa^{2} \gamma^{-2}  f(\tilde{v}) \tilde{v} \Delta N_{\mathrm{sust}}$ where $\Delta N_{\mathrm{sust}}$ is a
series of sustained slices small in the sense that $\langle \Delta \tilde{v} \rangle \ll \tilde{v}$.  In the large $\gamma$ limit, one may replace
$\langle \Delta \tilde{v} \rangle$ and $\Delta N_{\mathrm{sust}}$ with corresponding differential quantities, and in this regime
one has
\begin{equation}
f(\tilde{v}) = - \frac{\gamma^{2}}{\kappa^{2} \tilde{v} } \left( \frac{d \tilde{v}}{d N_{\mathrm{sust}}} \right)_{\mathrm{steady~state}}
\label{eq:Eq500}
\end{equation}
which may be extracted numerically from data obtained in the context of the relative area scenario for $\gamma \gg 1$ (i.e. for $\gamma = 2000$ in
this work).

The characteristic function $f(\tilde{v})$ in~Eq.\ref{eq:Eq500} in principle allows one access of  $N_{\mathrm{sust}}(\tilde{v})$
for a broad range of erosion scenarios distinct from the steady state case by exploiting the fact that
$d \tilde{v} = - f(\tilde{v}) \tilde{v}^{-1/3} A_{\mathrm{char}}(\tilde{v})^{2}  d N_{\mathrm{sust}}$
Solving for $d N_{\mathrm{sust}}$ and integrating yields
\begin{equation}
N_{\mathrm{sust}}(\tilde{v}) = \int_{\tilde{v}}^{1} \tilde{v}^{1/3} \frac{d \tilde{v}}{A_{\mathrm{char}}^{2}f(\tilde{v})}
\label{eq:Eq800}
\end{equation}

\begin{figure}
\includegraphics[width=.45\textwidth]{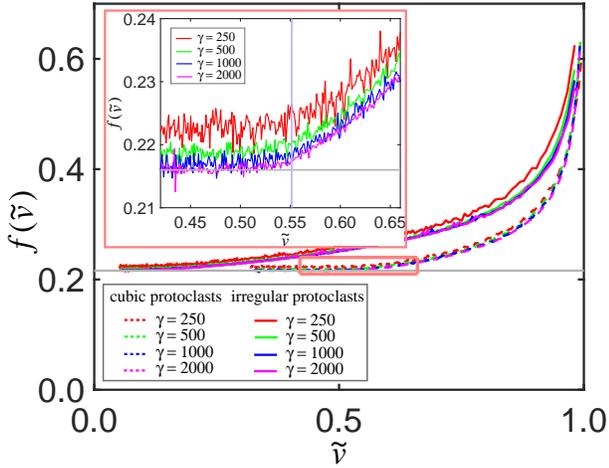}
\caption{\label{fig:Fig25} (Color online) Characteristic function $f(\tilde{v})$ for cube shaped protoclasts (broken lines) and
irregular protoclasts (solid lines).  the inset is a magnified view of the regions boxed in pink for cubic protoclasts.}
\end{figure}

Characteristic functions for cubic and irregular protoclasts are shown in Fig.~\ref{fig:Fig25}, with solid lines representing the former and 
broken lines the latter.  The inset graph is a magnified view of the region indicated by the pink box in the 
main graph for cubic protoclasts.  Common to both cases is the decline from an elevated value for $\tilde{v}$ near unity at the beginning of the erosion 
process, with $f(\tilde{v})$ leveling out at a finite value for $\tilde{v} \ll 1$ (i.e. for later times after a large portion 
of the original volume has eroded).  This asymptotic value, indicated in the main graph and inset plots with a solid gray line,
is 0.216(1) for both cube shaped and irregular protoclasts.
The elevated portion of $f(\tilde{v})$ for small times is compatible with an initially  accelerated loss of volume 
as corners and edges give way to regions of high local curvature which shed volume comparatively rapidly.

Although $f(\tilde{v})$ for the case of irregular and cube shaped counterparts appears to converge to a common 
asymptotic value, the characteristic function for the case of cubic protoclasts is distinct in abruptly reaching 0.216 for a finite 
volume fraction $\tilde{v} = 0.5732$, where the second order phase transition to shapes lacking primordial facets occurs.  
The singular behavior is highlighted in the inset, which shows a magnified view of $f(\tilde{v})$ for cube shaped protoclasts
with a slope discontinuity signaling the attainment of the long time value, and coinciding with the loss of primordial facets 
transitions indicated with the solid vertical line.

The flattening of $f(\tilde{v})$ following an initial rapid decrease provides an avenue for 
obtaining a closed form expression for time scales for the attainment of specific structural milestones at 
particular volume fractions $\tilde{v}$; given the 
structure of $f(\tilde{v})$, the relationship we obtain serves as an approximation as well as a rigorous upper bound for the time needed to
wear stones down to a particular volume fraction.  

\begin{figure}
\includegraphics[width=.45\textwidth]{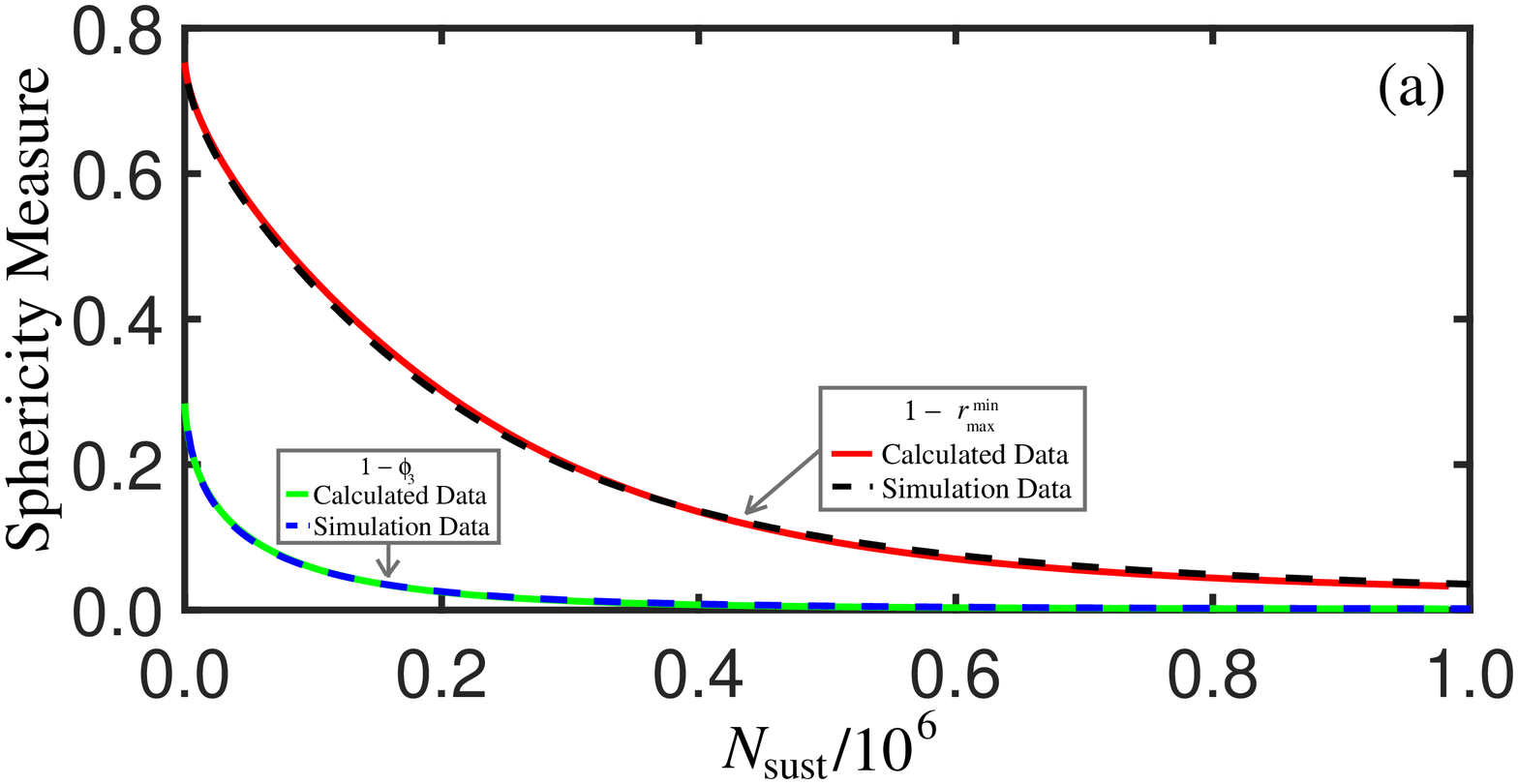}
\includegraphics[width=.45\textwidth]{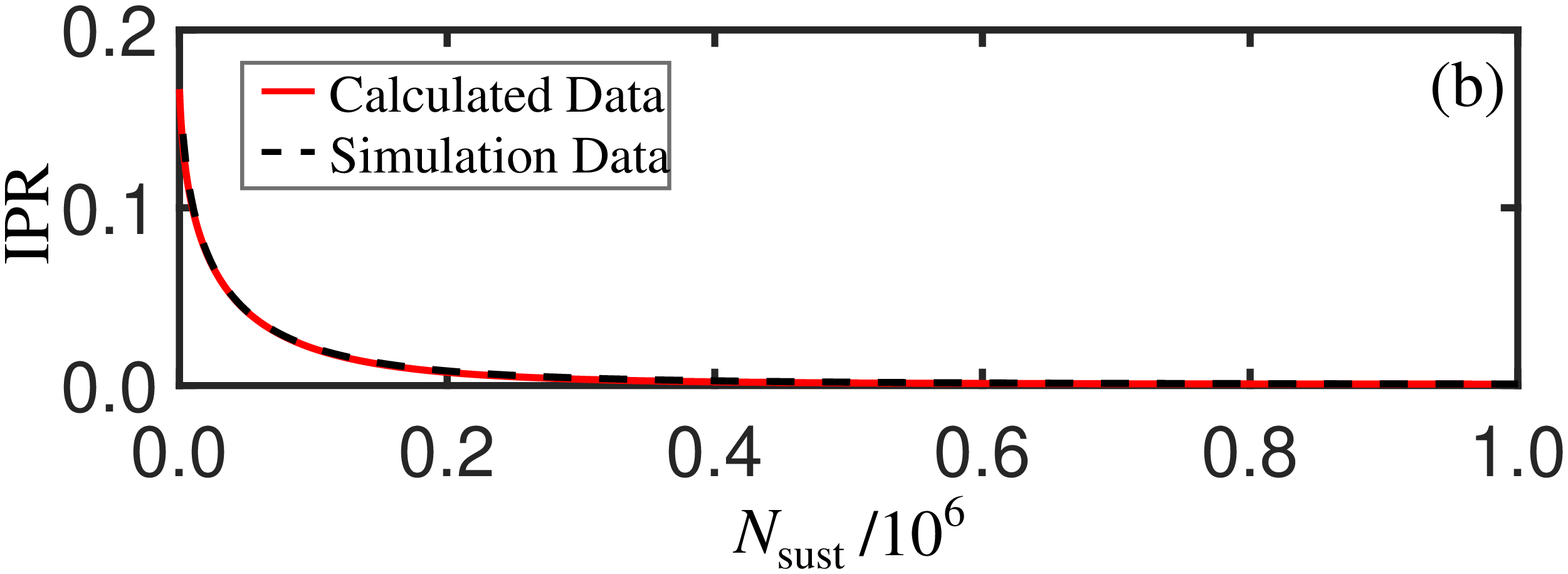}
\includegraphics[width=.45\textwidth]{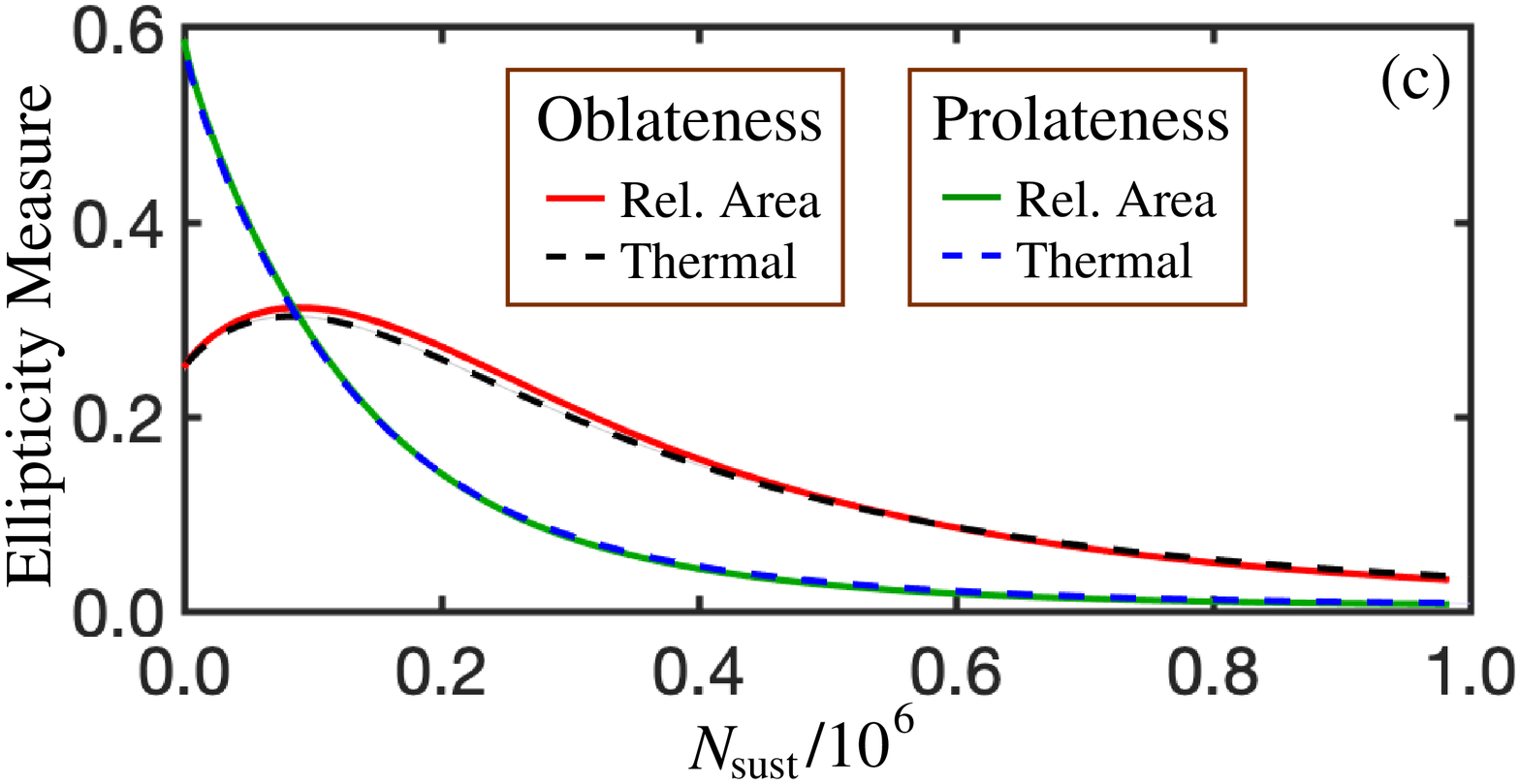}
\includegraphics[width=.45\textwidth]{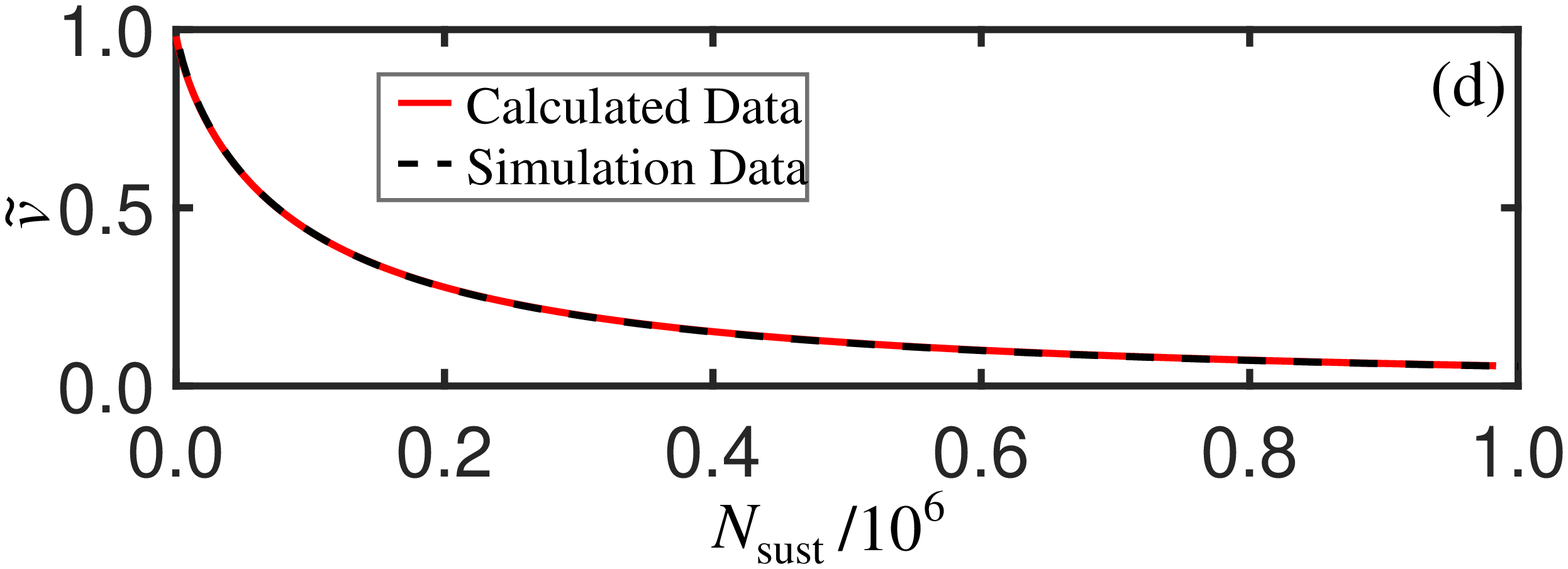}
\includegraphics[width=.45\textwidth]{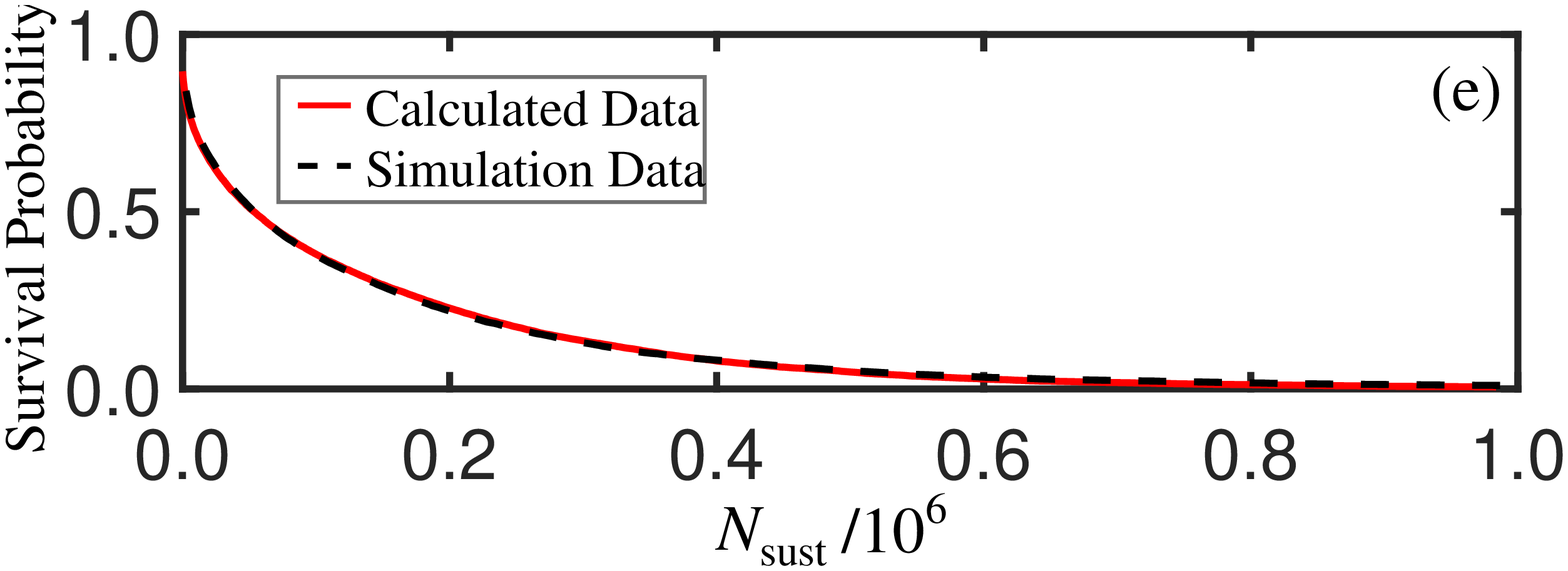}
\caption{\label{fig:Fig26} (Color online) Primordial facet survival probabilities are show in panel (a)  and 
panel (b) respectively;
non-steady state simulation data (broken lines) for $\gamma^{'} = 160$
and  calculated curves based on $\gamma = 2000$ steady state results (solid line) are plotted versus $N_{\mathrm{sust}}$ with
corresponding semilogarithmic graphs in the insets of panel (a) and panel (b).}
\end{figure}

Curves predicted based on relative area results (solid lines) and results from direct Monte 
Carlo calculations (broken traces) appear in panel (a) and panel (b) of Fig.~\ref{fig:Fig25}, respectively
with log-log graphs in the inset.  Broadly there is excellent agreement among the predicted and directly calculated
variables.  

\section{Salient Time Scales}

Due to the universal dependence of relevant variables on volume fraction $\tilde{v}$, to find time scales for a salient 
event such as a structural phase transition, one need only determine the volume fraction (e.g. in the context of the 
relative area scenario) for the case of interest and calculate the elapsed time in terms of sustained slices using
Equation~\ref{eq:Eq800}.  Since $f(\tilde{v})$ decreases monotonically, ultimately converting to the 
asymptotic value $f_{0} \equiv 0.216$, one has $f(\tilde{v}) \leq f_{0}$ and thus time scales gleaned from Equation~\ref{eq:Eq800} 
are approximately given and bounded from above by $f_{0} \int_{\tilde{v}}^{1} A_{\mathrm{char}}^{2} \tilde{v}^{\frac{1}{3}} d \tilde{v}$,
often amenable to exact calculation.  For an power law scaling of the kinetic energy where $A_{\mathrm{char}} = (\kappa /\gamma) \tilde{v}^{\alpha}$,
we obtain
\begin{equation}
\tilde{\tau} = \frac{3}{\kappa^{2}f_{0}(6 \alpha - 4)} [\tilde{v}^{4/3 - 2 \alpha} - 1]
\label{eq:Eq900}
\end{equation}

\begin{figure}
\includegraphics[width=.45\textwidth]{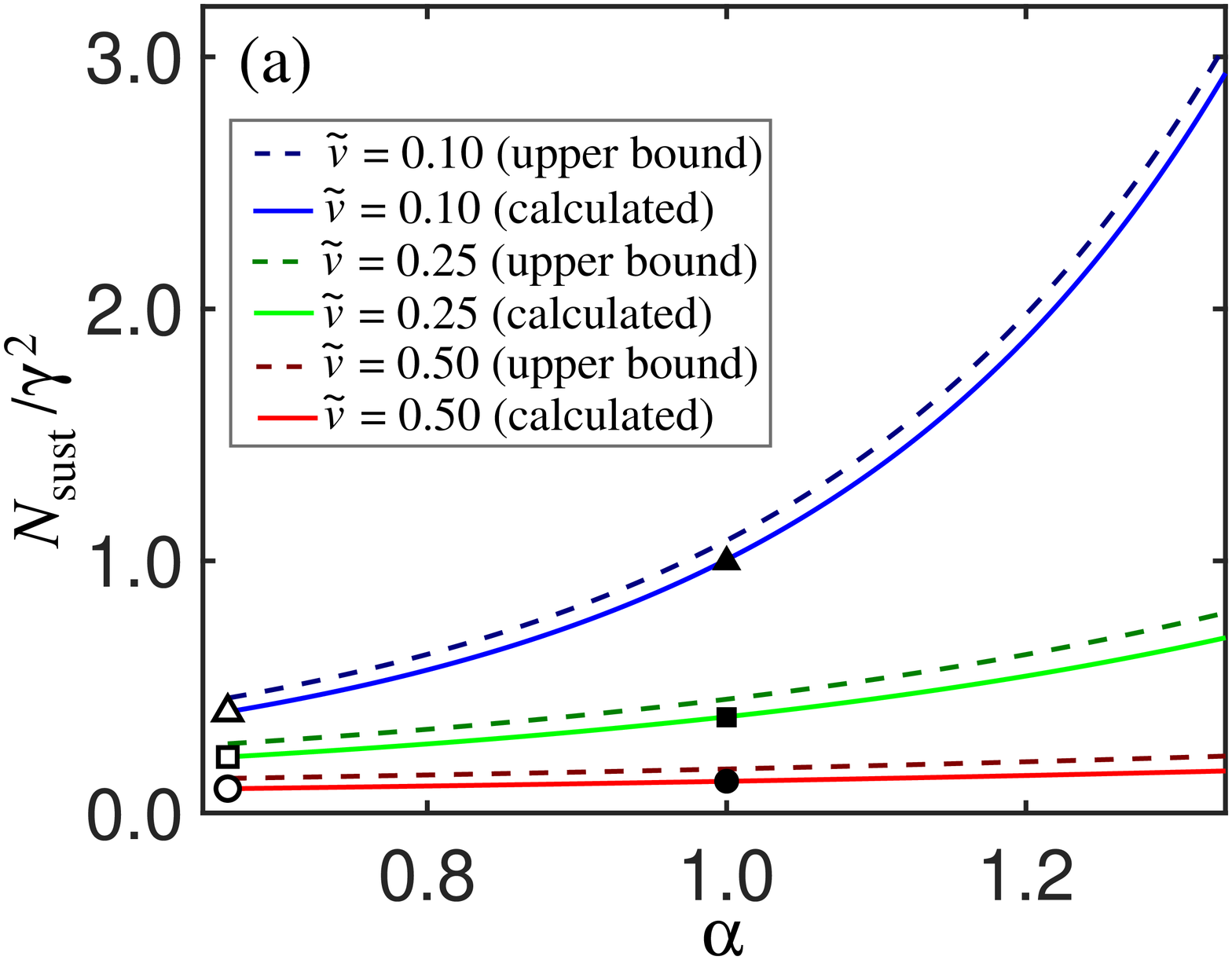}
\includegraphics[width=.35\textwidth]{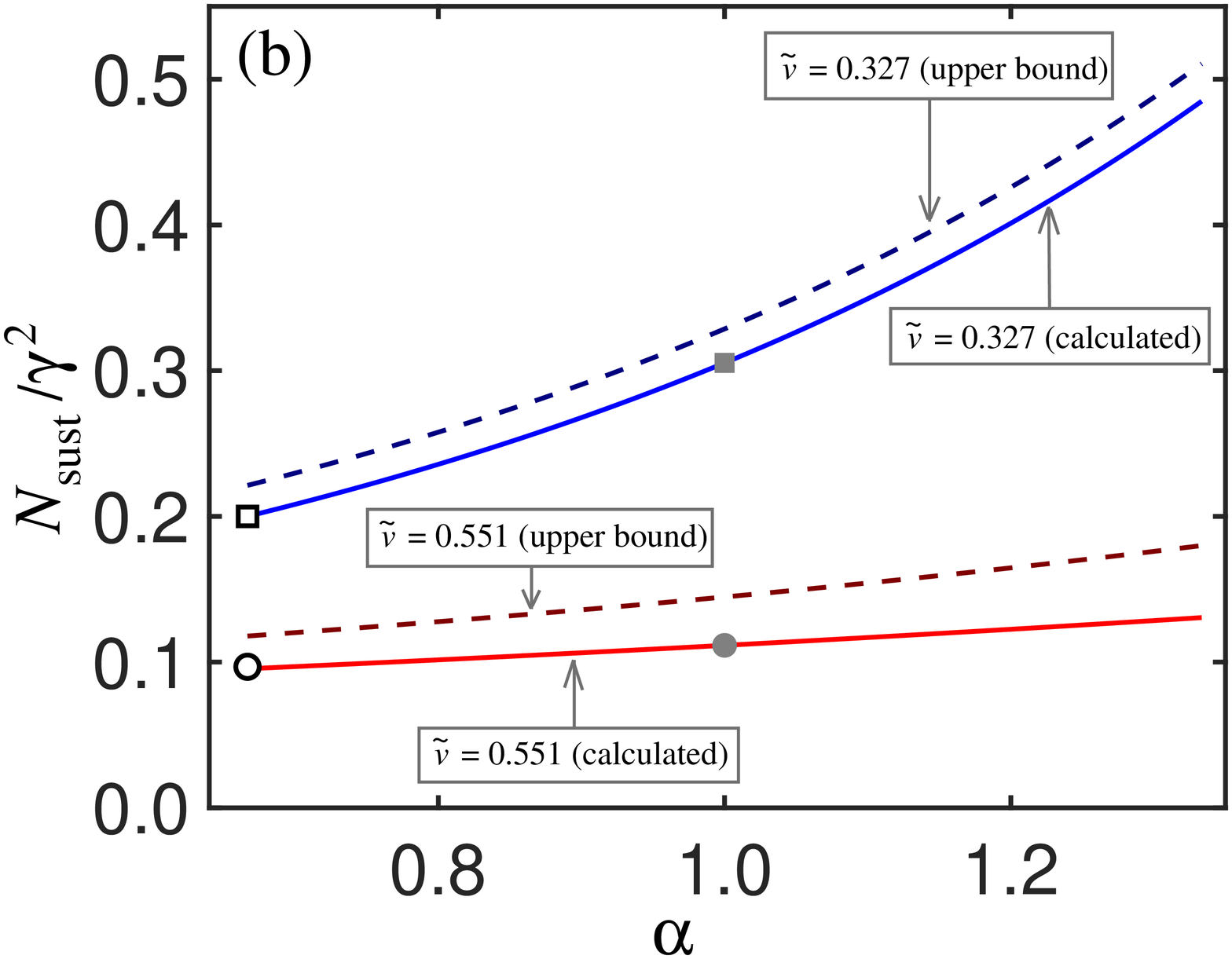}
\caption{\label{fig:Fig27} (Color online) Time scales for selected volume fractions $\tilde{v}$ for cubic and irregular 
protoclasts in panel (a) and panel (b) respectively.  Solid lines are calculated from direct Monte Carlo simulation results, while 
broken lines are closed form results bounding time scales from above. Open and filled symbols represent
direct simulation results for the relative area and fixed velocity scenarios, respectively.}
\end{figure}

Fig.~\ref{fig:Fig27} shows salient reduced time scales $\tilde{\tau}$ for a range of values of the exponent $\alpha$; solid curves are obtained 
from Monte Carlo simulation results in the case of the relative area scenario, while broken lines are an upper bound for time scales from Equation~\ref{eq:Eq900}. 
In panel (a) of Fig.~\ref{fig:Fig27}, pertaining to cube shaped protoclasts, the two time scales shown include
the structural phase transition involving the loss of primordial facets as well
as the attainment of spherical shapes (corresponding to equilibration in the relative area scenario) for volume fractions $\tilde{v} = 0.57$ and $\tilde{v} = 0.37$, 
respectively.  On the other hand, in panel (b) of Fig.~\ref{fig:Fig27}, for cohorts of irregular protoclasts, the three time scales shown correspond to 
$\tilde{v} = 0.50$, $\tilde{v} = 0.25$, and $\tilde{v} = 0.10$.  In spite of the monotonic rise in $\tilde{\tau}$ for the attainment of various volume
fraction milestones, panel (a) and panel (b) of Fig.~\ref{fig:Fig24} do not diverge for a finite value of $\alpha$.  Moreover, the upper bound in
Eq. (i.e. the same upper bound for the cases of cubic and irregular protoclasts), albeit increasing with increasing $\alpha$, 
nevertheless  remains finite for finite $\alpha$.

\section{Conclusions}

In conclusion, with large scale Monte Carlo simulations, we have examined the erosion of rocks through stochastic chipping, considering 
polyhedral stones with as many as 3,600 facets and averaging over at least 1000 realizations of 
disorder.  Using an energy based criterion for accepting a randomly chosen slicing plane,
our calculation is unique in placing no restrictions on the number of vertices and faces sheared away with each fracture event.
We have argued on theoretical grounds and verified by direct simulation that time scales for the removal of a 
specific amount of volume or the attainment of a particular structural milestone scale quadratically in the toughness parameter $\gamma$ 
which specifies the amount of energy per area associated with a fracture; the largest time scales examined in our calculations
exceed a million sustained slices per erosion sequence.

Cohorts of stochastically chipped protoclasts in the form of regular polyhedra undergo a structural phase transition in which all
primordial facets are abruptly lost with concomitant singularities in other relevant observables, marking a genuine second order   
phase transition; in a subsequent structural transformation evident in measures of sphericity, the stones revert to spherical shapes.  
More broadly, however, ensemble averaged observables in the case of initially identical irregular shapes 
are punctuated by an elimination of facets in multiple stages instead of a single event.  On the other hand, 
cohorts of irregular 
protoclasts exhibit no structural phase transitions as an aggregate, with individual eliminations of primordial surfaces being 
blurred by structural disorder.

We find direct measures of deviation from a spherical profile (i.e. the sphericity $\phi_{3}$ and ratio of minimum and maximum center 
of mass distances $r^{\mathrm{min}}_{\mathrm{max}}$) decrease monotonically with accumulating sustained slices.  
However, the oblateness measure actually rises at the initial stages of the erosion trajectory, reaching a peak when half of the 
original volume has been chipped away and then decreasing. This non-monotonic behavior of the oblateness is attributed to initially blade-like 
protoclasts which lose material more rapidly from the ends than from the edges, briefly becoming more oblate in shape before eventually being 
rounded into a spherical profile. 

That data from distinct erosion scenarios (i.e. relative area and fixed velocity schemes) collapse onto common curves when plotted with respect to 
the remaining volume fraction confirms the universal dependence of observables on the latter~\cite{Domokos1}.  This phenomenon 
provides an avenue for using results in the relative area case to calculate time dependent observables for arbitrary erosion schemes,
a technique we validate by reproducing results gleaned in the context of the fixed velocity scenario by direct Monte Carlo calculation. 
In addition, one also is afforded an efficient approach for calculating characteristic times (e.g. for specific remaining volume 
fractions) for a range of erosion scenarios.  
We have obtained closed form approximate expressions which bound these time scales from above, and which are in good quantitative 
agreement with the latter.


\begin{acknowledgments}
We acknowledge helpful discussions with Michael Crescimanno.
Calculations in this work have benefited from use of the Ohio Supercomputer facility (OSC)~\cite{OSC}.
\end{acknowledgments}


\end{document}